\documentclass[
  aps,
  prd,
  twocolumn,
  superscriptaddress,
  amsfonts,
  amssymb,
  amsmath,
  nofootinbib,
  longbibliography,
  preprintnumbers
]{revtex4-2}

\usepackage[T1]{fontenc}
\usepackage[utf8]{inputenc}
\usepackage{amsmath,amssymb,mathtools,mathrsfs}
\usepackage{graphicx}
\usepackage{booktabs,longtable,multirow,array}
\usepackage[usenames,dvipsnames]{xcolor}
\usepackage[unicode=true,pdfusetitle,bookmarks=true,
  bookmarksnumbered=true,bookmarksopen=true,breaklinks=true,
  pdfborder={0 0 0},colorlinks=true]{hyperref}
\hypersetup{citecolor=blue,filecolor=blue,linkcolor=blue,urlcolor=blue}

\allowdisplaybreaks

\newcommand{\dd}{\mathrm{d}}
\newcommand{\ii}{\mathrm{i}}

\newcommand{\Order}{\mathcal O}
\newcommand{\kdiss}{k_2^{(1)}}
\newcommand{\kstatic}{k_2^{(0)}}
\newcommand{\lambdadiss}{\lambda_2^{(1)}}
\newcommand{\lambdastatic}{\lambda_2^{(0)}}
\newcommand{\be}{\begin{equation}}
\newcommand{\ee}{\end{equation}}

\newcommand{\sone}{\beta_1}
\newcommand{\stwo}{\beta_2}
\newcommand{\sthree}{\beta_3}
\newcommand{\mathcalS}{\mathcal S}

\begin{document}

\title{Dissipative Tidal Love Numbers of Neutron Stars from Viscous Heating}

\author{Nikola Savi\'c}
\affiliation{Universit\'e Paris-Saclay, CNRS, CEA, Institut de Physique
Th\'eorique, 91191 Gif-sur-Yvette, France}
\author{Filippo Vernizzi}
\affiliation{Universit\'e Paris-Saclay, CNRS, CEA, Institut de Physique
Th\'eorique, 91191 Gif-sur-Yvette, France}

\date{\today}

\begin{abstract}
We compute the leading low-frequency dissipative tidal response of
nonrotating relativistic neutron stars.  We formulate the response in a
worldline effective field theory. We determine the linear dissipative Love
number in two complementary ways: from the redshifted viscous entropy
production inside the star and from the imaginary, frequency-odd part of
the exterior metric.  The first method reduces the calculation to an
integral over the static equilibrium tide; the second provides an
independent check.  We apply our approach to publicly available equations
of state and assess the corresponding correction to the gravitational-wave
phase of neutron-star binaries.  For an equal-mass $1.4+1.4M_\odot$ binary
with both stars at a redshifted temperature $T_\infty=10^5\,\mathrm K$ and a
standard electron shear viscosity, tidal dissipation shifts the
gravitational-wave phase by less than $10^{-3}$ cycles between
$20\,{\rm Hz}$ and $1\,{\rm kHz}$ for the four equations of state
considered.  Including the resulting rise in the stars' temperature in a
simple thermal model reduces the shift below $10^{-6}$ cycles.  For the
viscosity model considered here, shear-viscous tidal dissipation in cold
neutron stars is negligible for gravitational-wave phasing. This suggests
that a measurable dissipative tidal signature would have to come from a
different mechanism.
\end{abstract}

\maketitle
\tableofcontents

\section{Introduction}
\label{sec:introduction}

Neutron stars provide a unique laboratory for matter at densities several
times larger than nuclear saturation density, where the equation of state
remains poorly constrained by terrestrial experiments. The equation of
state determines the macroscopic properties of a star, including its radius,
maximum mass, and response to external perturbations. Gravitational-wave
observations of binary neutron-star inspirals have opened a complementary
way of probing this regime, with the tidal information contained in
GW170817 already providing constraints on neutron-star radii and dense-matter
models
\cite{LIGOScientific:2017vwq,LIGOScientific:2018hze,
Chatziioannou:2020pqz}.

At large separations, the dynamics of a compact binary is determined
primarily by the masses and spins of its constituents. As the separation decreases, the tidal field generated by each
body deforms its companion and makes the orbital evolution sensitive to
internal structure
\cite{Flanagan:2007ix,Hinderer:2007mb,Damour:2009vw,
Hinderer:2009ca}. 
These finite-size effects leave a cumulative imprint on the
gravitational-wave phase.

For a nonrotating object, the leading tidal response is the mass
quadrupole induced at linear order by the gravitoelectric tidal field in
the static limit. Its amplitude is characterized by the quadrupolar Love
number $\kstatic$
\cite{Hinderer:2007mb,Binnington:2009bb,Damour:2009vw}.

 Conservative tidal effects have been extended in two complementary
directions. At linear order in the tidal field, the leading frequency-dependent
conservative correction appears at order $\omega^2$ and describes
dynamical tides
\cite{Pitre:2023xsr,HegadeKR:2024agt,Katagiri:2024wbg}. Recent work has matched this response to a worldline EFT and computed
its effects on gravitational waveforms~\cite{Apostolidis:2026qsg}
and on the binary binding energy~\cite{Saketh:2026bcx}. At second order in the tidal field, the response is characterized by
quadratic Love numbers, introduced in Ref.~\cite{Poisson:2020vap}. In the static limit, these coefficients vanish for Schwarzschild black
holes~\cite{Poisson:2020vap,Riva:2023rcm,Iteanu:2024dvx}, a property that
can be understood in terms of symmetries and extended to all nonlinear
orders~\cite{Combaluzier-Szteinsznaider:2024sgb,Kehagias:2024rtz,
Parra-Martinez:2025bcu}. For neutron stars, Pitre and Poisson computed the
quadrupolar quadratic Love number for relativistic polytropic models and
assessed its impact on dynamical tides \cite{Pitre:2025qdf}. The static
quadratic response was subsequently calculated for realistic equations of
state in Ref.~\cite{Pani:2025qxs}, together with its contributions to the
binary dynamics and gravitational waveform.

In this paper we focus instead on the leading dissipative response. At
frequencies well below the characteristic internal frequencies of the
object, the retarded quadrupole can be expanded as
\begin{equation}
 Q^{ij}(\omega)
 =
 \frac{R_*^5}{3G}
 \left[
 \kstatic
 +\ii\omega R_*\kdiss
 +\Order(\omega^2R_*^2)
 \right]
 E^{ij}(\omega) \;.
 \label{eq:introresponse}
\end{equation}
The normalization is chosen so that the coefficients in brackets are
dimensionless. The frequency-odd coefficient $\kdiss$ determines the
out-of-phase quadrupole and the leading irreversible tidal heating
\cite{Ripley:2023qxo,HegadeKR:2024agt,Saketh:2024juq}. For black holes it describes absorption by the horizon, and tidal-heating
parameters of this type have been constrained with binary-black-hole
observations~\cite{Chia:2024bwt}, whereas for neutron stars it is
controlled by transport processes in the stellar fluid.

In a binary, the energy absorbed by the individual bodies contributes to
the balance law in addition to the flux radiated to infinity. For
nonspinning objects, the resulting correction to the gravitational-wave
phase first enters at relative 4PN order
\cite{Ripley:2023qxo,Saketh:2024juq}. Although this is one PN order earlier
than the static quadrupolar tidal correction, its magnitude
is controlled by the efficiency of the microscopic dissipation and must be
determined by matching the long-distance response to the internal dynamics
of the compact object.

In this paper we determine $\kdiss$ for nonrotating relativistic neutron
stars by matching the worldline effective theory to stellar perturbation
theory in two complementary ways. First, we derive a covariant relation
between local viscous entropy production and the increase of the
gravitational mass measured at infinity. At leading order in frequency and
viscosity, matching this mass-growth rate to the worldline prediction
expresses $\kdiss$ as a manifestly positive radial integral involving only
the static, inviscid equilibrium tide. Second, we solve directly for the
frequency-odd metric perturbation and extract the coefficient of its
decaying exterior solution, which is then matched to the same worldline
response. The two methods are shown to be equivalent. 
We then use these results to evaluate $\kdiss$ for a set of publicly available
equations of state and find values of order $\mathcal O(10^{-5})$, leading to a shift of less than $10^{-3}$ gravitational-wave cycles between   $20\,{\rm Hz}$ and
$1\,{\rm kHz}$.

A complementary determination of the neutron-star dissipation number,
based on gravitational Raman scattering and relativistic stellar
perturbation theory, was presented in Ref.~\cite{Saketh:2024juq}.
During the completion of this work, a unit-conversion error in the
viscosity used in that analysis was corrected (see the \texttt{v4} arXiv
version of Ref.~\cite{Saketh:2024juq}).  The corrected dissipative Love
numbers and their contribution to the gravitational-wave phase agree with
our results when evaluated on the same stellar background and with the
same viscosity profile.

The paper is organized as follows.  Section~\ref{sec:response} fixes the
EFT response and normalization conventions.  Section~\ref{sec:heating}
derives the relativistic mass-growth law.  In Sec.~\ref{sec:linearLove} we
obtain $\kdiss$ from entropy production and from the frequency-odd metric,
and show that the two expressions are equivalent.
Section~\ref{sec:numerics} presents the numerical results for four representative EOSs,
Sec.~\ref{sec:waveform} derives the leading waveform correction and
estimates the effect of the heating of the stars during the inspiral, and
Sec.~\ref{sec:conclusions} concludes.  Moreover, App.~\ref{app:SK} derives the
retarded response in the Schwinger--Keldysh formalism,
App.~\ref{app:metricmatching} gives the metric calculation and the proof of
equivalence,  and App.~\ref{app:heating} describes the effect of tidal heating on the star temperature.

We use the mostly-plus signature for the metric, $(-, +, +, +)$ and units $c=1$ unless it is restored explicitly. Throughout we use Greek letters $\mu,\nu, \ldots$ for indices of the four-tensors, lowercase Latin letters $i,j,\ldots$ for indices of spatial tensors and capital Latin letters $A, B, C,\ldots,$ to denote angular indices on the two-dimensional   sphere $S^2$. Symmetric and the symmetric trace-free (STF) combinations of indices are denoted by $A_{(\mu}B_{\nu)}$ and $A_{\langle\mu}B_{\nu \rangle}$ respectively.
Our Fourier convention is
\begin{equation}
 f(t)=\int\frac{\dd\omega}{2\pi}\,
 e^{-\ii\omega t}\,\widetilde f(\omega) \;.
 \label{eq:fourier}
\end{equation}

\section{Dissipative tidal response in the worldline EFT}
\label{sec:response}

Dissipative finite-size effects have a natural formulation in the worldline
effective field theory of compact objects~\cite{Goldberger:2004jt,
Goldberger:2005cd,Goldberger:2007hy,Porto:2016pyg}.
At distances much larger than the stellar
radius $R_*$, the EFT action can be organized as
\begin{equation}
 S=S_{\rm EH}+S_{\rm pp}+S_{\rm tid} \;,
 \label{eq:fullaction}
\end{equation}
where
\begin{equation}
 S_{\rm EH}=\frac{1}{16\pi G}\int\dd^4x\sqrt{-g} R
\end{equation}
is the Einstein-Hilbert action and
\begin{equation}
 S_{\rm pp}=-m \int\dd\tau \;
 \label{eq:EHpp}
\end{equation}
describes the minimal coupling of the object to gravity.
Here $m $ is the mass parameter measured with respect to the asymptotic time
and $\tau$ is the proper time along the object worldline,
$\dd\tau = \sqrt{- g_{\mu \nu} \dd x^\mu \dd x^\nu}$. Denoting by $z(\tau)$ the worldline of the compact object, the object four-velocity is given by $U^\mu = \dd z^\mu/\dd\tau$, with $U_\mu U^\mu = -1$.

Finite-size effects
are encoded in $S_{\rm tid}$. 
At leading order in the field and derivatives, for parity-even modes, this is given by
\begin{equation}
 S_{\rm tid}=\int\dd\tau\,Q^{ij} (\tau) E_{ij} (z (\tau) )\;,
 \label{eq:QEcoupling}
\end{equation}
where $Q^{ij}$ is the induced quadrupole of the object, related to the Newtonian STF mass quadrupole by $I^{ij}=-2Q^{ij}$ (see App.~\ref{sec:equivalence}), while $E_{ij}$ is the electric part of the Weyl tensor projected on an
orthonormal spatial triad $e_i^\mu$ orthogonal to $U^\mu$,
\begin{equation}
 E_{ij}=C_{\mu\alpha\nu\beta}
 U^\mu e_i^\alpha U^\nu e_j^\beta \;.
 \label{eq:electricWeyl}
\end{equation}

The finite-size response of the compact object is encoded in the dynamics of the quadrupole $Q_{ij}$. For the conservative response and for frequencies smaller than the fundamental modes of the star, $Q_{ij} $  can be integrated out and the effective action written  in terms of non-minimal gravitational operators localized on the point particle's worldline, organized by the number of fields and (spatial and time) derivatives \cite{Goldberger:2005cd,Goldberger:2020fot}. Similarly, one can derive a local action for the dissipative response. However, as tidal dissipation transfers orbital energy into the internal degrees of freedom of the compact object---viscous fluid motion for neutron stars or  horizon absorption for black holes---a dissipative response cannot be derived from a conventional (single-branch) action. 

As reviewed in App.~\ref{app:SK}, to compute the response in the presence of dissipation one has to resort to the in-in, or Schwinger-Keldysh (SK), formalism \cite{Schwinger:1960qe,Keldysh:1964ud,Galley:2012hx,Galley:2012qs}, in which all fields are doubled, $E_{ij}\to E^\pm_{ij}$, corresponding to the two branches of the closed time contour.  
In this case, the effective action can be expressed in terms of the Keldysh combinations
\be
E^c_{ij} = \frac{E^+_{ij} + E^-_{ij}}{2} \;, \qquad E^r_{ij} = E^+_{ij} - E^-_{ij} \;,
\ee
respectively the classical  and the response fields.
Integrating out the worldline multipoles defines the SK generating functional, whose part linear in the response field controls the mean induced quadrupole, while the remainder encodes noise and fluctuations and  does not contribute to the mean response.
Here we will only need the former. 

Expanding the retarded response kernels at low frequency and keeping terms
quadratic in the tidal field, the deterministic part of the generating
functional defines the effective SK action
(see Eq.~\eqref{eq:appWRlocal} in App.~\ref{app:SK})
\begin{equation}
 S_{\rm eff}
  =
  \int \dd\tau\, E_r^{ij}
  \Big[
2\lambda^{(0)}_{2}\,E^c_{ij}
- 2\lambda^{(1)}_{2}\,\dot E^c_{ij}  +  \ldots
  \Big]
  \;,
  \label{eq:WRsplitQ}
\end{equation}
where overdots denote derivatives with respect to the proper time $\tau$ along the worldline, $\dot E_{ij} = U^\mu \nabla_\mu E_{ij}$ and the ellipses denote terms higher order in the time derivatives.
Here $\lambda^{(0)}_{2}$ is the Wilson coefficient of the conservative action---the linear static tidal Love number---while $\lambda^{(1)}_{2}$ is the leading dissipative (linear-in-frequency) coefficient, arising from the odd-in-$\omega$, imaginary part of the retarded response
kernel [see Eq.~\eqref{eq:appGRexpansion}].
A passive object, such as a neutron star, can absorb energy but cannot spontaneously supply it. In this case, the coefficient controlling leading-order tidal dissipation must be non-negative $\lambda^{(1)}_{2}\ge0$ (see App.~\ref{app:SK}).

The induced deterministic quadrupole is obtained by varying
$S_{\rm eff}$ once with respect to the response field $E_r$ and then
setting $E_r$ to zero~\cite{Liu:2018kfw},
\be
Q^{ij}(\tau)   = \left. \frac{\delta S_{\rm eff}}{\delta E^r_{ij} } \right|_{E_r=0, \; E_c = E} \;.
\label{eq:QfromW}
\ee
This gives
\begin{equation}
 Q^{ij}(\tau)
 =2\lambdastatic E^{ij}(\tau)
  -2\lambdadiss\dot E^{ij}(\tau) + \ldots
  \;,
 \label{eq:QresponseTime}
\end{equation}
which in Fourier space becomes
\be
 Q^{ij}(\omega)=2\lambda_2(\omega)E^{ij}(\omega) \;,
\ee
with
\be
 \lambda_2(\omega) =\lambdastatic
       +\ii\omega\lambdadiss+\Order(\omega^2) \;.
 \label{eq:lambdaresponse}
\ee
Reality implies $\lambda_2(-\omega)=\lambda_2(\omega)^*$: the static
coefficient is the leading real, frequency-even response, whereas
$\lambdadiss$ is the leading imaginary, frequency-odd response.

It is convenient to define a dimensionless frequency-dependent Love number
by
\begin{equation}
 k_2(\omega)\equiv\frac{6G}{R_*^5}\lambda_2(\omega)
 =\kstatic+\ii\omega R_*\kdiss+\Order(\omega^2R_*^2),
 \label{eq:k2frequency}
\end{equation}
so that
\begin{equation}
 \kstatic=\frac{6G}{R_*^5}\lambdastatic,
 \qquad
 \kdiss=\frac{6G}{R_*^6}\lambdadiss.
 \label{eq:k2definitions}
\end{equation}
 In these conventions
$\kstatic$ agrees with the standard quadrupolar gravitoelectric Love number
\cite{Hinderer:2007mb}.\footnote{Comparison with the response convention of
Ref.~\cite{Saketh:2024juq} gives $\kdiss=\nu_2^E$. Thus, their
$H_\omega^E =\tfrac23\nu_2^E
 \left( {R_*}/{Gm}\right)^6 $ is related to our coefficient by
$ \lambdadiss=\tfrac14m(Gm)^5H_\omega^E $. }

The interaction Hamiltonian associated with
Eq.~\eqref{eq:QEcoupling} is
$H_{\rm tid}=-Q^{ij}E_{ij}$. Therefore, the instantaneous power supplied by
the time-dependent external tidal field to the internal degrees of freedom
is
\begin{equation}
 P_{\rm tid}
 \equiv \frac{\partial H_{\rm tid}}{\partial\tau}
 =-Q^{ij}\dot E_{ij}.
 \label{eq:tidalpower}
\end{equation}
Using Eq.~\eqref{eq:QresponseTime}, this power separates into
conservative and dissipative contributions,

\begin{align}
 -Q^{ij}\dot E_{ij}
 &=
 -2\lambdastatic E^{ij}\dot E_{ij}
 +2\lambdadiss\dot E^{ij}\dot E_{ij}
 +\ldots
 \nonumber\\
 &=
 -\frac{\dd}{\dd\tau}
 \left(
 \lambdastatic E^{ij}E_{ij}
 \right)
 +2\lambdadiss\dot E^{ij}\dot E_{ij}
 +\ldots .
 \label{eq:tidalworkdecomposition}
\end{align}
The first term is a total derivative; it describes the reversible exchange of conservative tidal energy and can  be absorbed into the definition of the conservative energy
of the tidally deformed system.
Its integrated contribution depends only on the initial and final tidal
configurations and, in particular, vanishes over a closed cycle.
The remaining term is the irreversible power absorbed
by the internal degrees of freedom. In the  rest frame of the object, the absorbed internal energy
contributes to its rest mass,
\begin{equation}
 \dot m_{\rm EFT}
 =
 2\lambdadiss\dot E^{ij}\dot E_{ij}.
 \label{eq:EFTheating}
\end{equation}
We will later match this expression to the  mass increase due to tidal heating from the internal fluid dynamics, which we compute next.

\section{Relativistic viscous heating}
\label{sec:heating}

Away from equilibrium, the star is described by an imperfect fluid with stress-energy tensor 
\begin{equation}
 T_{\mu\nu}
 =\rho u_\mu u_\nu+p h_{\mu\nu}
  -\zeta\Theta h_{\mu\nu}-2\eta\sigma_{\mu\nu} \;.
 \label{eq:viscousstress}
\end{equation}
Here $\rho$ and $p$ are respectively the
perfect fluid  energy density  and
pressure, $u^\mu$ denotes the four-velocity of the fluid element and we have introduced the spatial projector
$ h_{\mu\nu}=g_{\mu\nu}+u_\mu u_\nu $.
Dissipation enters through the last two terms on the right-hand side of the above equation,
proportional respectively to the bulk viscosity $\zeta$ and the shear
viscosity $\eta$. We have defined the expansion scalar of the fluid congruence by
\begin{equation}
 \Theta\equiv\nabla_\mu u^\mu,
\end{equation}
and the shear tensor by
\begin{equation}
 \sigma_{\mu\nu}
 \equiv
 h_\mu{}^\alpha h_\nu{}^\beta\nabla_{(\alpha}u_{\beta)}
 -\frac13h_{\mu\nu}\Theta .
 \label{eq:kinematics}
\end{equation}The latter is STF and orthogonal to the flow,
$\sigma_{\mu\nu}u^\mu=0$.

Let   $n$ be the baryon number density. Baryon number conservation reads
\begin{equation}
u^\mu \partial_\mu n + \Theta n =0 \;.
 \label{eq:baryoncons}
\end{equation}
Projecting $\nabla_\mu T^\mu{}_\nu=0$ along $u^\nu$ gives~\cite{Langlois:2006iq}
\begin{align}
u^\mu \partial_\mu \rho+\Theta (\rho+p) =\mathcal D \equiv
 2\eta\sigma_{\mu\nu}\sigma^{\mu\nu}+\zeta\Theta^2 \;.
  \label{eq:localenergy}
\end{align}
Using these two equations in  the local
first law of thermodynamics,    
\begin{equation}
 \dd\rho=\frac{\rho+p}{n}\dd n+nT\dd s \;,
 \label{eq:localfirstlaw}
\end{equation}
with $s$ the entropy per baryon,  the reversible
compression work cancels, leaving 
a simple expression for the entropy density variation,
\begin{equation}
 nT u^\mu \partial_\mu s=\mathcal D \;.
 \label{eq:entropyproduction}
\end{equation}
Thus $\mathcal D$ is the heat generated per unit proper volume and per unit
fluid proper time.
The local second law requires $\mathcal D\geq0$. In the fluid rest
frame, $\sigma_{\mu\nu}\sigma^{\mu\nu}$ is a sum of squares. Since
shear and expansion can vary independently, nonnegative entropy
production for arbitrary flows requires $\eta\geq0$ and $\zeta\geq0$.

Equation~\eqref{eq:entropyproduction} is local.  To compare it with the
worldline heating rate \eqref{eq:EFTheating}, the locally deposited heat must
be converted into the gravitational mass measured at infinity.  To do so, we introduce the redshift factor 
\begin{equation}
 \chi\equiv- (\partial_t)^\mu u_\mu \;,
 \label{eq:redshiftdefinition}
\end{equation}
where $(\partial_t)^\mu$ is the hypersurface-orthogonal Killing
vector of the unperturbed star, normalized to unit norm at infinity.
On the static background, $u^\mu=(\partial_t)^\mu / \chi$ and therefore
\begin{equation}
 u^t=\frac1\chi \;.
 \label{eq:clockredshift}
\end{equation}

We treat the slowly driven star as a quasi-static sequence of
neighboring equilibria at fixed total baryon number, and assume that
there is no matter or entropy flux through its surface.  The required
global relation is then the relativistic first law for neighboring
stellar equilibria.  At fixed total baryon number and circulation, the
irreversible entropy contribution to the variation of the
gravitational mass is~\cite{Friedman:1988er,Friedman:2001pf,Green:2013ica}
\begin{equation}
 \delta m
 =\int_\Sigma\dd V\,\chi nT\,\delta s \;,
 \label{eq:stellarfirstlaw}
\end{equation}
where the integral is performed on the static
$t={\rm const}$ hypersurface $\Sigma$, and $\dd V$ is its proper three-volume element.
The factor of $\chi$ in Eq.~\eqref{eq:stellarfirstlaw} has a direct
physical interpretation: local heat
$\delta Q=nT\,\delta s$ deposited into a fluid element contributes
$\chi\delta Q$ to the energy defined with respect to the Killing time at
infinity.
Taking the rate of Eq.~\eqref{eq:stellarfirstlaw} for comoving fluid elements
and using Eqs.~\eqref{eq:entropyproduction} and
\eqref{eq:clockredshift}, we obtain
\begin{align}
 \frac{\dd m}{\dd t}
 =
 \int_\Sigma\dd V\,\chi^2
 \left(
 2\eta\sigma_{\mu\nu}\sigma^{\mu\nu}
 +\zeta\Theta^2
 \right).
 \label{eq:massgrowthtarget}
\end{align}
One factor of $\chi$ redshifts the locally deposited energy to infinity,
as in Eq.~\eqref{eq:stellarfirstlaw}, while the additional factor converts
the rate per unit fluid proper time into a rate per unit asymptotic time.  At the perturbative order considered
here, the redshift factor and hypersurface measure may be evaluated on
the unperturbed static background: their tidal corrections multiply a
heating rate that is already quadratic in the perturbations and
therefore contribute only at higher order.  Likewise, the relative
Lorentz factor between the fluid and the static hypersurface differs
from unity only at quadratic order in the fluid velocity.

Equation~\eqref{eq:massgrowthtarget} gives the irreversible secular
increase of the gravitational mass. We  
compute this quantity and match it to  the EFT heating rate \eqref{eq:EFTheating} in the next section.

\section{The  dissipative Love number}
\label{sec:linearLove}

At leading order in frequency and viscosity, the heating can be evaluated
entirely on the inviscid equilibrium tide.  Indeed, the static tidal
displacement is linear in the tidal field, and its slow time dependence
induces a velocity and a shear of order $\omega E$.  The viscous heating rate
therefore scales as
\begin{equation}
 \eta\,\sigma_{\mu\nu}\sigma^{\mu\nu}
 =\Order(\eta\omega^2E^2) \;.
 \label{eq:ordersigmasq}
\end{equation}
Viscous corrections to the fluid motion
contribute only at higher order in $\eta$ while finite-frequency corrections to the inviscid displacement affect the
heating only at higher order in $\omega$.

\subsection{Background star and equilibrium tide}
\label{sec:equilibriumtide}

The static background metric can be written as
\begin{equation}
 \dd s^2=-e^{\Phi(r)}\dd t^2+e^{\Lambda(r)}\dd r^2
          +r^2\dd\Omega^2 \;.
 \label{eq:staticmetricpreview}
\end{equation}
We define the enclosed mass ${\cal M}(r)$
by
\begin{equation}
 e^{-\Lambda(r)}=1-\frac{2G {\cal M}(r)}{r} \;.
 \label{eq:backgrounddefinitions}
\end{equation}
With this definition, the background profiles satisfy the Tolman--Oppenheimer--Volkoff
equations~\cite{Tolman:1939jz,Oppenheimer:1939ne},
\begin{align}
 {\cal M}'&=4\pi r^2\bar\rho \;,
 \nonumber\\
 \bar p'&=-\frac{G(\bar\rho+\bar p)
  [{\cal M}+4\pi r^3\bar p]}
 {r^2(1-2G{\cal M}/r)} \;, 
 \nonumber\\
 \Phi'&=-\frac{2\bar p'}{\bar\rho+\bar p}\;.
 \label{eq:TOV}
\end{align}
We then close the system with a barotropic equilibrium equation of state
$\bar\rho=\bar\rho(\bar p)$.
Regularity fixes ${\cal M}(0)=0$, the surface is defined by
$\bar p(R_*)=0$, and exterior matching fixes
$e^{\Phi(R_*)}=1-2Gm/R_*$, where the total gravitational mass is defined by
\begin{equation}
 m\equiv {\cal M}(R_*) \;.
 \label{eq:totalmass}
\end{equation}

We work in Regge--Wheeler gauge~\cite{Martel:2005ir} and focus on slowly varying, even-parity quadrupolar perturbations,
\begin{equation}
\begin{split}
 \delta g_{\mu\nu}\dd x^\mu\dd x^\nu={}&-S\big[
 e^\Phi H_0(r)\dd t^2+e^\Lambda H_2(r)\dd r^2
 \\
 &+r^2K(r)\dd\Omega^2\big]
 -2H_1(r)\dot S\,\dd t\dd r \;.
\end{split}
 \label{eq:metricperturbation}
\end{equation}
The angular and time dependence is carried by
\begin{equation}
 S(t,\theta,\phi)\equiv E_{ij}(t)n^in^j \;,
 \qquad n^i\equiv\frac{x^i}{r} \;.
 \label{eq:angularscalar}
\end{equation}
Since $E_{ij}$ has
dimensions of inverse length squared, the radial functions $H_0,H_2,K$ have
dimensions of length squared. Regularity requires
$H_0,H_2,K\propto r^2$ near the center.

At equilibrium, the four velocity of the fluid element is  $
 \bar u^\mu=e^{-\Phi/2}\delta^\mu_t $.
We parametrize the Lagrangian displacement by
\begin{equation}
 \xi^r=e^{-\Lambda/2}\frac{W}{r}S,
 \qquad
 \xi^A=-\frac{V}{r^2}\gamma^{AB} D_B S \;,
 \label{eq:displacement}
\end{equation}
where $A,B$ are angular indices, $\gamma_{AB}$ is the metric on the unit
two-sphere and  $D_A$ is the covariant derivative of $\gamma_{AB}$.  To the required order, the four-velocity perturbation is
\begin{equation}
 \delta u^t=-\frac12e^{-\Phi/2}H_0S \;,
 \qquad
 \delta u^i=e^{-\Phi/2}\dot\xi^i \;.
 \label{eq:velocityperturbation}
\end{equation}

At order $\omega^0$, the angular projection of
$\delta(\nabla_\mu T^\mu{}_{\nu})=0$ gives
\begin{equation}
 \delta p=-\frac{\bar p+\bar\rho}{2}H_0S \;.
 \label{eq:angularEuler}
\end{equation}
Substitution into the radial projection of the same conservation equation
gives
\begin{equation}
 \delta\rho=-\frac{\bar p+\bar\rho}{2}
 \frac{\bar\rho'}{\bar p'}H_0S
 =\frac{\bar\rho'}{\Phi'}H_0S \;,
 \label{eq:fluidperturbations}
\end{equation}
where the second equality follows from Eq.~\eqref{eq:TOV}.  

The trace-free angular Einstein equation equates $H_2-H_0$ to the
anisotropic stress.  This stress vanishes for the inviscid perfect-fluid
equilibrium tide, while its first viscous contribution is of order
$\eta\omega$.  We may therefore set $H_0=H_2\equiv H$ at the order considered
here.  The mixed radial--angular Einstein equation then gives
\begin{equation}
 K'=H'+\Phi'H \;,
 \label{eq:Kprime}
\end{equation}
whereas a scalar combination of the radial and angular Einstein equations,
after use of Eq.~\eqref{eq:TOV}, gives the algebraic constraint
\begin{align}
 K={}&\left[1-2\pi Gr^2(3\bar p+\bar\rho)
 +e^{-\Lambda}\frac{r\Phi'}2
       \left(1+\frac{r\Phi'}2\right)\right]H
 \nonumber\\
 &+\frac{r^2\Phi'}4e^{-\Lambda}H' \;.
 \label{eq:Kalgebraic}
\end{align}
Differentiating Eq.~\eqref{eq:Kalgebraic} and imposing
Eq.~\eqref{eq:Kprime} eliminates $K$ and gives the standard static tidal
equation~\cite{Hinderer:2007mb},
\begin{equation}
\begin{split}
 H''+\bigg\{\frac2r+e^\Lambda\bigg[
 \frac{2G {\cal M}}{r^2}+4\pi Gr(\bar p-\bar\rho)\bigg]\bigg\}H'
 \\
 +\bigg\{4\pi G e^\Lambda\bigg[
 5\bar\rho+9\bar p+\frac{\bar p+\bar\rho}{c_{\rm eq}^2}\bigg]
 -\frac{6e^\Lambda}{r^2}-(\Phi')^2\bigg\}H=0 \;,
\end{split}
 \label{eq:Hmaster}
\end{equation}
where $c_{\rm eq}^2=\dd\bar p/\dd\bar\rho$.  Regularity at the center selects
the interior solution
\begin{equation}
 H_{\rm int}(r)=\mathcal C r^2+\Order(r^4) \;,
 \label{eq:Hcentralbehavior}
\end{equation}
where $\mathcal C$ is arbitrary because Eq.~\eqref{eq:Hmaster} is
homogeneous.

We integrate the regular solution to the stellar surface and match it to
the general static vacuum solution,
\begin{equation}
 H_{\rm ext}(r)=A_\uparrow H_\uparrow(r)
             +A_\downarrow H_\downarrow(r) \;.
 \label{eq:exteriordecomposition}
\end{equation}
For the quadrupolar perturbation, we choose the vacuum
basis~\cite{Hinderer:2007mb}
\begin{equation}
 H_\uparrow(r)=r(r-r_s) \;,
 \qquad
 H_\downarrow(r)=\frac{5}{r_s^3}\,
 Q_2^2\!\left(\frac{2r}{r_s}-1\right) ,
 \label{eq:exteriorbasis}
\end{equation}
with $r_s\equiv2Gm$. Here $Q_2^2(x)$ is the
associated Legendre function of the second kind for $x \equiv 2r/r_s-1 >1$,
\begin{equation}
 Q_2^2(x)=\frac32\,(x^2-1)\ln\frac{x+1}{x-1}
 -\frac{x(3x^2-5)}{x^2-1} \;.
 \label{eq:Q22}
\end{equation}
The normalization is chosen so that
$H_\downarrow=r^{-3}\big[1+\tfrac32 r_s/r+\Order(r_s^2/r^2)\big]$.
Correspondingly, $H_\uparrow=(Gm)^2(x^2-1)$ is proportional to $P_2^2(x)$.

For the EOS considered here, the surface junction conditions reduce to
continuity of $H$ and $H'$, namely
\begin{align}
 H_{\rm int}(R_*)&=A_\uparrow H_\uparrow(R_*)
        +A_\downarrow H_\downarrow(R_*) \;,
 \nonumber\\
 H_{\rm int}'(R_*)&=A_\uparrow H_\uparrow'(R_*)
        +A_\downarrow H_\downarrow'(R_*) \;.
 \label{eq:staticmatching}
\end{align}
Because $S$  in Eq.~\eqref{eq:metricperturbation} already contains the physical tidal amplitude, we
can set $A_\uparrow=1$.  The two matching conditions then determine the central
normalization $\mathcal C$ and the response amplitude $A_\downarrow$.

The decaying amplitude determines the standard static Love number. The result is conveniently 
expressed in terms of the compactness and the
logarithmic surface derivative,
\begin{equation}
 C\equiv\frac{Gm}{R_*} \;,
 \qquad
 y_0\equiv\frac{R_*H'(R_*)}{H(R_*)} \;.
 \label{eq:ycompactness}
\end{equation}
One then obtains $\kstatic=F(C,y_0)$, where~\cite{Hinderer:2007mb}
\begin{align}
 F(C,y)&\equiv\frac{8C^5}{5}(1-2C)^2\frac{A(C,y)}{D(C,y)} \;,
 \label{eq:Hindererk2}\\
 A(C,y)&=2+2C(y-1)-y \;,
 \nonumber\\
 D(C,y)&=2C[6-3y+3C(5y-8)]
 \nonumber\\
 &\quad+4C^3[13-11y+C(3y-2)+2C^2(1+y)]
 \nonumber\\
 &\quad+3(1-2C)^2A(C,y)\ln(1-2C) \;.
 \nonumber
\end{align}

The static field equations determine $H$ and the static Love number,
while leaving the fluid displacement $\xi^\mu$ undetermined. This is needed to compute
the velocity and shear. We consider a stratified star in which each
fluid element moves adiabatically and retains its composition over a
tidal cycle.
This description requires the tide to vary slowly compared with the gravity
modes ($g$-modes) that the composition gradient of the core
supports~\cite{Reisenegger:1992,Lai:1993di,Yu:2016ltf}: only then can the
fluid elements adjust so that $\Delta p=\Delta\rho=0$, as shown below. For
faster tides the flow no longer preserves the volume of each fluid element,
and both the shear and the bulk-viscous contributions to the heating
change. The $g$-mode frequencies of cold neutron stars are, however,
estimated to lie between about $100\,{\rm Hz}$ and a few hundred
hertz~\cite{Reisenegger:1992,Lai:1993di,Yu:2016ltf}, so this condition
holds only at the low-frequency end of the inspiral band. The fixed-temperature waveform estimates of Sec.~\ref{sec:waveform},
which are dominated by the highest frequencies, should therefore
be regarded as an extrapolation.

For any background scalar $\bar q(r)$, the Lagrangian perturbation is
\begin{equation}
 \Delta q\equiv\delta q+\mathcal L_\xi\bar q
 =\delta q+\xi^r\bar q'.
 \label{eq:Lagrangianperturbation}
\end{equation}
Equations~\eqref{eq:angularEuler} and~\eqref{eq:fluidperturbations}
give $\delta p=c_{\rm eq}^2\delta\rho$, and the same relation holds
for the Lagrangian perturbations because
$\bar p'=c_{\rm eq}^2\bar\rho'$. The adiabatic equation of state at
fixed composition also gives $\Delta p=c_{\rm ad}^2\Delta\rho$,
where $c_{\rm ad}$ is the sound speed at fixed entropy per baryon
and composition. In a stratified region, $c_{\rm ad}^2\ne c_{\rm eq}^2$,
so these two relations imply $\Delta p=\Delta\rho=0$ at leading order.
Since $c_{\rm ad}$ does not appear in the result, cold barotropic EOS
tables suffice to evaluate it.

At fixed entropy per baryon and composition, the first law gives
$\Delta\rho=(\bar\rho+\bar p)\Delta n/\bar n$, so
$\Delta n=0$~\cite{Saketh:2024juq}. Since each fluid element conserves
its baryon number, its proper volume is also unchanged at this order.
Then, using Eqs.~\eqref{eq:angularEuler} and \eqref{eq:displacement}, the condition
$\Delta p=0$ reads
\begin{equation}
 0=\Delta p
 =\left[-\frac{\bar p+\bar\rho}{2}H
 +e^{-\Lambda/2}\frac{\bar p'}{r}W\right]S,
 \label{eq:Deltapzero}
\end{equation}
and determines the radial displacement.  Linearized baryon conservation
gives
\begin{equation}
 \frac{\Delta n}{\bar n}
 =\left[\frac{H+2K}{2}
 -\frac{e^{-\Lambda/2}}{r^2}(rW'+W)
 -\frac{6V}{r^2}\right]S \;.
 \label{eq:Deltan}
\end{equation}
Imposing $\Delta n=0$ therefore determines the angular displacement.  The
two profiles are
\begin{align}
 W&=\frac{re^{\Lambda/2}(\bar p+\bar\rho)}{2\bar p'}H \;,
 \label{eq:Wprofile}\\
 V&=\frac1{12}\left[r^2(H+2K)
 -2e^{-\Lambda/2}(rW'+W)\right].
 \label{eq:Vprofile}
\end{align}

The metric profiles $H$ and $K$ and the displacement profiles $W$ and $V$
provide all the ingredients needed to evaluate the heating integral \eqref{eq:massgrowthtarget} at
leading order.

\subsection{Entropy-production integral}
\label{sec:entropymethod}

We evaluate the heating on a uniformly rotating quadrupolar tide with constant amplitude $\mathcal E$,
\begin{equation}
 E_{ij}(t)=\mathcal E\,
 n^{\rm o}_{\langle i}(t)n^{\rm o}_{j\rangle}(t) \;,
 \qquad
 n_{\rm o}^i=(\cos\Omega t,\sin\Omega t,0) \;,
 \label{eq:rotatingtide}
\end{equation}
where angular brackets denote the STF projection.  In the rest frame of the star, its proper time coincides with the
asymptotic Killing time $t$, so that
$\dot E_{ij}\dot E^{ij}=2\mathcal E^2\Omega^2$ and Eq.~\eqref{eq:EFTheating} becomes 
\begin{equation}
\frac{\dd m_{\rm EFT}}{\dd t} =4\lambdadiss\mathcal E^2\Omega^2.
 \label{eq:rotatingEFTheating}
\end{equation}
Matching consists of comparing the  right-hand side of this equation
with the right-hand side of Eq.~\eqref{eq:massgrowthtarget}.

The calculation of the latter is simplified by
three properties of the equilibrium tide.  
First, at
linear order the expansion   is
\begin{equation}
 \Theta=e^{-\Phi/2}\left[
 \frac{e^{-\Lambda/2}}{r^2}(rW'+W)+\frac{6V}{r^2}
 -\frac{H+2K}{2}\right]\dot S \;,
 \label{eq:expansionintermediate}
\end{equation}
where the  brackets contain the fractional rate of change of the proper
volume of a comoving fluid element.  Substituting
Eq.~\eqref{eq:Vprofile} gives
\begin{equation}
 \Theta=0\;,
 \label{eq:vanishingexpansion}
\end{equation}
i.e., the leading equilibrium-tide flow preserves the proper volume of each
fluid element, so bulk viscosity, which would contribute for a compressible
tidal flow~\cite{HegadeKR:2024agt}, does not enter the leading tidal heating.
Second, because  of Eq.~\eqref{eq:ordersigmasq}, we can evaluate  the radial-angular integration
weight in Eq.~\eqref{eq:massgrowthtarget} on the {\em background} metric \eqref{eq:staticmetricpreview}. This gives
\begin{equation}
 \chi^2\dd V
 =r^2e^{\Phi+\Lambda/2}\dd r\dd\Omega \;,
 \label{eq:heatingweight}
\end{equation}
where we have used  $\chi=e^{\Phi/2}$ and
$\dd V=e^{\Lambda/2}r^2\dd r\dd\Omega$. 
Third, although of 
$O(\omega)$, $H_1$ drops out of the linear shear and does not contribute to the entropy-production integral at
leading order. Consequently, the leading
$O(\omega^2)$ heating depends only on $H$, $K$, $W$, and $V$, computed above.

The computation is most transparent 
when projecting the shear onto an
orthonormal frame comoving with the fluid.  Since
$\sigma_{\mu\nu}u^\mu=0$, its temporal components vanish in this frame.
At the order considered, the frame can be evaluated on the static
background: the radial leg is the unit vector $e^{-\Lambda/2}\partial_r$,
and the angular legs are orthonormal vectors tangent to the sphere of
radius $r$.  Angular components below refer to these legs, with $D_A$ and
$\gamma_{AB}$ the covariant derivative and metric on the unit sphere.
The nonvanishing components are
\begin{subequations}
\label{eq:shearcomponents}
\begin{align}
 \sigma_{r r }&=-\frac{e^{-(\Phi+\Lambda)/2}}{3r^2}  \sone\dot S,
 \\
 \sigma_{r  A}&=-\frac{e^{-(\Phi+\Lambda)/2}}{2r^2} \stwo D_A\dot S,
 \\
 \sigma_{AB}&=\frac{e^{-(\Phi+\Lambda)/2}}{r^2}\left[
 \frac{\sone}{6}\gamma_{AB}\dot S
 -\sthree\dot S_{\langle AB\rangle}\right] \;,
\end{align}
\end{subequations}
where
we have introduced
the symmetric trace-free tensor
\begin{equation}
 \dot S_{\langle AB\rangle}
 \equiv D_AD_B\dot S- \tfrac12 \gamma_{AB} D^2 \dot S \;,
 \label{eq:sheartensorharmonic}
\end{equation}
and the radial
amplitudes
\begin{subequations}
\label{eq:shearamplitudes}
\begin{align}
 \sone={}&e^{\Lambda/2}r^2(H-K)+6e^{\Lambda/2}V
        -2rW'+4W \;,
 \\
 \stwo={}&rV'-2V-e^{\Lambda/2}W \;,
 \\
 \sthree={}&e^{\Lambda/2}V \;.
\end{align}
\end{subequations}
Contracting with the background metric yields
\begin{equation}
\begin{split}
 \sigma_{\mu\nu}\sigma^{\mu\nu}
 =\frac{e^{-\Phi-\Lambda}}{r^4}\left[
 \frac{\sone^2}{6}\dot S^2
 +\frac{\stwo^2}{2}D_A\dot S D^A\dot S\right.
 \\
 \left.{}+\sthree^2\dot S_{\langle AB\rangle}
          \dot S^{\langle AB\rangle}\right].
\end{split}
 \label{eq:shearsquaredpointwise}
\end{equation}

The above expression can be integrated on the  sphere using that,
for Eq.~\eqref{eq:rotatingtide}, the angular scalar introduced in
Eq.~\eqref{eq:angularscalar} is
\begin{equation}
 S=\mathcal E\left[
 \sin^2\theta\cos^2(\phi-\Omega t)- 1/3\right].
 \label{eq:Srotating}
\end{equation}
This gives
\begin{equation}
 \int\dd\Omega\,\sigma_{\mu\nu}\sigma^{\mu\nu}
 =\frac{8\pi}{45}\frac{\Omega^2\mathcal E^2}{r^4}
 e^{-\Phi-\Lambda}\mathcal B(r) \;,
 \label{eq:shearsquaredangular}
\end{equation}
where
\begin{equation}
 \mathcal B(r)=\sone^2+18 \stwo^2+72 \sthree^2 \;,
 \label{eq:positivekernel}
\end{equation}
and we have used that $ \int\dd\Omega\,\dot S^2
 =\tfrac{16\pi}{15} \Omega^2 \mathcal E^2$.\footnote{This follows either by direct integration or from the STF identity
\begin{equation}
 \int\dd\Omega\,
 (A_{ij}n^in^j)(B_{kl}n^kn^l)
 =\frac{8\pi}{15}A_{ij}B^{ij} \;.
 \label{eq:STFangularidentity}
\end{equation}}
Note that $\mathcal B(r)$ is manifestly non-negative, implying $\dot m \ge 0$ as expected.
Using the measure \eqref{eq:heatingweight}, Eq.~\eqref{eq:massgrowthtarget}
now becomes
\begin{equation}
 \frac{\dd m}{\dd t}
 =\frac{16\pi}{45}\mathcal E^2\Omega^2
 \int_0^{R_*}\dd r\,
 \frac{\eta(r)e^{-\Lambda/2}}{r^2}\mathcal B(r).
 \label{eq:stellarheatingradial}
\end{equation}
Matching this expression to Eq.~\eqref{eq:rotatingEFTheating} and using Eq.~\eqref{eq:k2definitions} yields
\begin{equation}
 \kdiss =\frac{8\pi G}{15 R_*^6}\int_0^{R_*}\dd r\,
 \frac{\eta(r)e^{-\Lambda/2}}{r^2}\mathcal B(r) \;.
 \label{eq:k2dissintegral}
\end{equation}
This is the main result of the entropy-production method.  It involves
only the static inviscid tide, is linear in the viscosity, and makes
passivity $\kdiss\geq0$ manifest.  It also does not require the
frequency-odd metric perturbation: the matching is performed on the
absorbed power, a gauge-invariant quantity, and the only metric matching
needed is the static one, which normalizes the tidal amplitude.

As a check, consider a uniform-density, incompressible Newtonian star
with constant shear viscosity $\eta$. In the weak-viscosity,
low-frequency regime $\nu_{\rm kin}/R_*^2\ll|\omega|\ll(Gm/R_*^3)^{1/2}$,
where $\nu_{\rm kin}=\eta/\bar\rho$, the heating integral gives
\begin{equation}
 \kdiss=\frac{25\pi}{2}\frac{\eta R_*^3}{Gm^2} \;.
 \label{eq:newtoniank2diss}
\end{equation}
This agrees with the coreless limit of Eq.~(49) of
Ref.~\cite{Ogilvie:2009qz}, after converting to our Love-number convention.

\subsection{Metric response}
\label{sec:metricmethod}

The dissipative Love number can also be extracted by solving the
metric perturbation equations through linear order in $\omega$ and
matching the interior solution to the exterior vacuum solution.
The interior calculation parallels that of Ref.~\cite{Saketh:2024juq},
while the matching follows the same approach as the dynamical-response
calculation of Ref.~\cite{HegadeKR:2024agt}.

We write
\begin{equation}
 H_0(r,\omega)
 =H(r)+\ii\omega R_*h(r)
 +\Order(\omega^2R_*^2) \;,
 \label{eq:oddmetricexpansionmain}
\end{equation}
where $H$ is the static equilibrium-tide solution.  Note that,  at order
$\eta\omega$, $H_0$ is different from $H_2$.  At linear order in
the viscosity, $h$ satisfies an inhomogeneous equation whose
homogeneous part is the static tidal equation and whose source is
constructed entirely from the equilibrium-tide profiles.  The complete
interior equations, source terms, and regular central solution are
given in App.~\ref{app:metricmatching}.

The vacuum perturbation equations contain no term linear in $\omega$.
Consequently, in the buffer region
$R_*\ll r\ll\omega^{-1}$, the frequency-odd exterior solution is a
linear combination of the same growing and decaying solutions used in
the static calculation.  It can therefore be matched through the
complex surface logarithmic derivative
\begin{align}
 y(\omega)&\equiv
 \left.\frac{rH_0'(r,\omega)}{H_0(r,\omega)}\right|_{R_*}
 =y_0+\ii\omega R_*y_1+\Order(\omega^2R_*^2) \;,
 \nonumber\\
 y_1&=
 \left.
 \frac{R_*\mathcal W[H,h]}{H^2}
 \right|_{R_*} \;,
 \label{eq:y1wronskian}
\end{align}
where $\mathcal W[H,h]\equiv Hh'-H'h$ is the Wronskian.
Note that the coefficient $y_1$ is invariant under a common rescaling of
$H$ and $h$ and under the addition of a multiple of $H$ to $h$.

Writing the static matching formula as
$k_2=F(C,y)$ and analytically continuing it to complex
$y(\omega)$\footnote{Reference~\cite{Saketh:2024juq} instead
parametrizes the surface data by
$T=T_0-\ii\omega R_*T_1+\Order(\omega^2R_*^2)$; see its
Eqs.~(5.7) and (5.8).  Its Eq.~(6.15) gives the algebraic relation
between $T_0$ and $y_0$, whose frequency expansion similarly relates
$T_1$ to $y_1$.}
gives
\begin{equation}
 \kdiss=F_{,y}(C,y_0)y_1 \;.
 \label{eq:k2metricresult}
\end{equation}
Thus the leading dissipative Love number can also be determined from
the imaginary, frequency-odd exterior metric using surface data alone.

Appendix~\ref{app:metricmatching} proves the equivalence of this result
with the entropy-production calculation.  A Green identity converts
the surface Wronskian in Eq.~\eqref{eq:y1wronskian} into
\begin{equation}
 F_{,y}(C,y_0)y_1
 =\frac{8\pi G}{15R_*^6}
 \int_0^{R_*}\dd r\,
 \frac{\eta e^{-\Lambda/2}}{r^2}\mathcal B(r) \;,
 \label{eq:methodsequivalent}
\end{equation}
which reproduces the entropy-production result~\eqref{eq:k2dissintegral}.

\section{Numerical results}
\label{sec:numerics}

We solve the TOV and tidal systems using cold, beta-equilibrated EOS tables
from the public CompOSE database~\cite{Typel:2013rza}.  Our main set consists
of APR(APR)~\cite{Akmal:1998cf}, RG(SLy4)~\cite{Gulminelli:2015csa},
PCP(BSk24)~\cite{Pearson:2018tkr}, and
GPPVA(DD2)~\cite{Typel:2009sy}.  These models span radii
$11.34\mbox{--}13.19\,{\rm km}$ at $1.4M_\odot$ while supporting maximum
masses above $2M_\odot$.

The evaluation of Eq.~\eqref{eq:k2dissintegral} requires a model for the
viscosity profile $\eta$ inside the star.  In neutron-star matter, shear
viscosity is controlled by microscopic transport processes that redistribute
momentum among the constituent particles.  It therefore depends on the local thermodynamic state,
 in particular on the density and temperature.  Although newly formed
neutron stars can be very hot, efficient cooling by neutrino emission~\cite{Yakovlev:2004iq}
is expected to reduce their internal temperature to
$T \sim10^5$--$10^6\,{\rm K}$ by the stage relevant for
late-inspiral binaries~\cite{Lai:1993di,Haskell:2012vg}.

We therefore adopt the following fitting formula as a benchmark,
following Refs.~\cite{Cutler:1987,Flowers:1976,Saketh:2024juq},
\begin{equation}
 \eta_{\rm cgs}(r)
 =6\times10^{26}
 \left(
   \frac{\rho_{\rm m} (r)}
        {10^{15}\,{\rm g\,cm^{-3}}}
 \right)^2
 \left(
   \frac{T(r)}{10^5\,{\rm K}}
 \right)^{-2}
\frac{\rm g}{\rm cm\, s} \;,
 \label{eq:viscositybenchmark}
\end{equation}
where $\rho_{\rm m}(r)$ is the local mass-equivalent density,\footnote{For the mass-equivalent density entering the viscosity
prescription we use $\rho_{\rm m}=\epsilon/c^2$, where $\epsilon$
is the total energy density supplied by the EOS.} expressed in
${\rm g\,cm^{-3}}$, and $T(r)$ is the local temperature in
kelvin.\footnote{This fit should not be regarded as  a precision model of composition-dependent and paired
neutron-star matter.  In particular, uncertainties associated with
particle fractions, effective masses, and superfluid gaps are not encoded
in the cold barotropic EOS tables employed here.}
In our $c=1$ convention, the viscosity entering
Eq.~\eqref{eq:k2dissintegral} is $\eta=\eta_{\rm cgs}/c$.
Thus $\eta$ has units of mass per area, and the combination
$G\eta R_*$ appearing in the dissipative Love number is dimensionless.

Moreover, we assume an isothermal star, i.e.,
that the 
redshifted (Tolman) temperature, 
\begin{equation}
 T_\infty\equiv e^{\Phi(r)/2}T(r) \;,
 \label{eq:Tolmantemperature}
\end{equation}
is uniform throughout the star \cite{Tolman:1930}.  
 The local temperature entering
Eq.~\eqref{eq:viscositybenchmark} is therefore $T(r)=  T_\infty e^{-\Phi(r)/2}$. In the numerical analysis below we  set
$T_\infty=10^5\,{\rm K}$. 
Since the dissipative Love number is linear in the viscosity we can relate the Love number computed at $T_\infty=10^5\,{\rm K}$  to that for other values of $T_\infty$ using
\begin{equation}
 \kdiss(T_\infty)
 =
 \kdiss(10^5\,{\rm K})
 \left(
   \frac{10^5\,{\rm K}}{T_\infty}
 \right)^2.
 \label{eq:k2temperaturescaling}
\end{equation}

Before evaluating Eq.~\eqref{eq:k2dissintegral} numerically, it is
useful to estimate the expected magnitude of the result.  From
Eqs.~\eqref{eq:QresponseTime} and \eqref{eq:k2definitions}, the
effective tidal lag time is \cite{HegadeKR:2024agt}
\begin{equation}
 \tau_{\rm lag}\equiv
 \frac{\lambdadiss}{\lambdastatic}
 =\frac{R_*}{c}\frac{\kdiss}{\kstatic} \;,
 \label{eq:lagtime}
\end{equation}
where $c$ has been restored.  Indeed, for a harmonic tide the ratio
between the out-of-phase and in-phase parts of the quadrupolar
response is $\omega\tau_{\rm lag}$.

Consider a fluid displacement of characteristic amplitude $\xi$
varying at angular frequency $\omega$.  Its velocity scales as
$v\sim\omega\xi$.  The viscous force density contains two spatial
derivatives of the velocity and therefore scales as
\begin{equation}
 f_\eta\sim
 \frac{\eta_{\rm cgs}\omega\xi}{R_*^2} \;.
\end{equation}
The characteristic stellar dynamical frequency satisfies
$\omega_*^2\sim Gm/R_*^3$, so the gravitational and pressure
restoring forces have the scale
\begin{equation}
 f_{\rm rest}\sim
 \rho_{\rm m}\omega_*^2\xi
 \sim\rho_{\rm m}\frac{Gm}{R_*^3}\xi \;.
\end{equation}
Identifying the ratio $f_\eta/f_{\rm rest} $ with $\omega \tau_{\rm lag} $, gives \cite{Chia:2024bwt}
\begin{equation}
\tau_{\rm lag} \sim \frac{\nu_{\rm kin} R_*}{G m} \;, \qquad
 \nu_{\rm kin}\equiv
 \frac{\eta_{\rm cgs}}{\rho_{\rm m}} \;, \label{eq:nukin}
\end{equation}
and using  Eq.~\eqref{eq:lagtime},
we find
\begin{equation}
 \kdiss\sim\kstatic\frac{c\nu_{\rm kin}}{Gm} \;.
 \label{eq:backofenvelope}
\end{equation}
At a characteristic mass-equivalent density
$\rho_{\rm m}= 10^{15}\,{\rm g\,cm^{-3}}$ and temperature
$T=10^5\,{\rm K}$, Eq.~\eqref{eq:viscositybenchmark} gives
 $\nu_{\rm kin}
 \sim6\times10^{11}\,{\rm cm^2\,s^{-1}}$.
For a $1.4M_\odot$ star,
$ {Gm}/{c}\simeq
 6.2\times10^{15}\,{\rm cm^2\,s^{-1}}$,
giving
\begin{equation}
 \frac{c\nu_{\rm kin}}{Gm}\sim10^{-4}.
\end{equation}
Since $\kstatic\sim0.1$, the dimensional estimate
\eqref{eq:backofenvelope} gives
$\kdiss\sim10^{-5}$, up to stellar-structure factors of order unity,
in agreement with the full calculation below.

For each EOS, we first construct the stable stellar sequence by varying
the central density.  The tabulated relation $\rho(p)$ is
interpolated in logarithmic variables and the TOV equations and the
static quadrupolar tidal equation are integrated simultaneously with a
fourth-order Runge--Kutta scheme.  We use a maximum radial step of
$0.01\,{\rm km}$, reducing the final step near the stellar surface.

At the surface, the numerical values of $H$ and $H'$ are matched to the
growing and decaying Schwarzschild exterior solutions.  This determines
both the static Love number and the rescaling required to set the
growing tidal amplitude to unity.  The corresponding profiles $K$, $W$,
and $V$ are then reconstructed from the static field equations, and
$\kdiss$ is obtained by direct numerical evaluation of
Eq.~\eqref{eq:k2dissintegral}.

\begin{figure*}[t]
 \centering
 \includegraphics[width=0.45\textwidth]
 {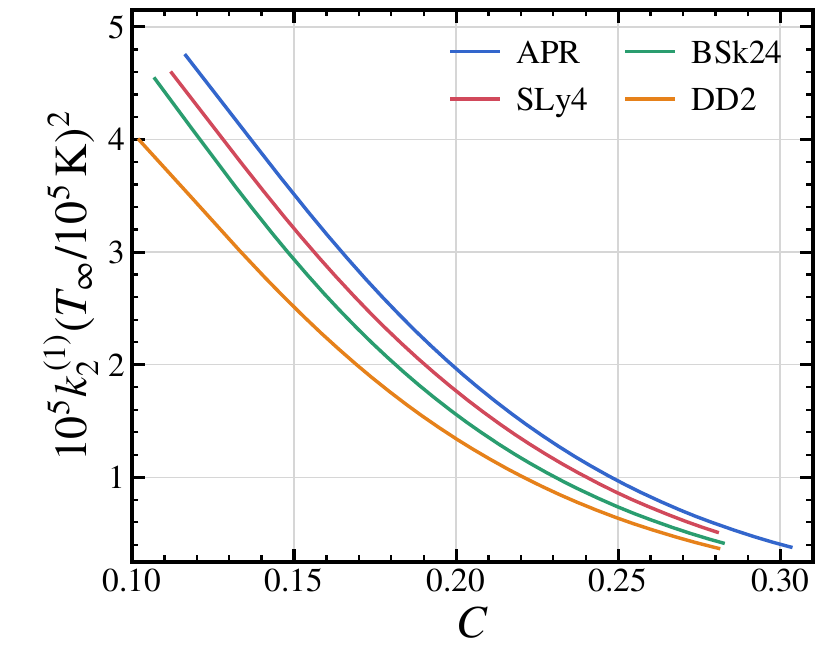}
 \hspace{1cm}
  \includegraphics[width=0.45\textwidth]
 {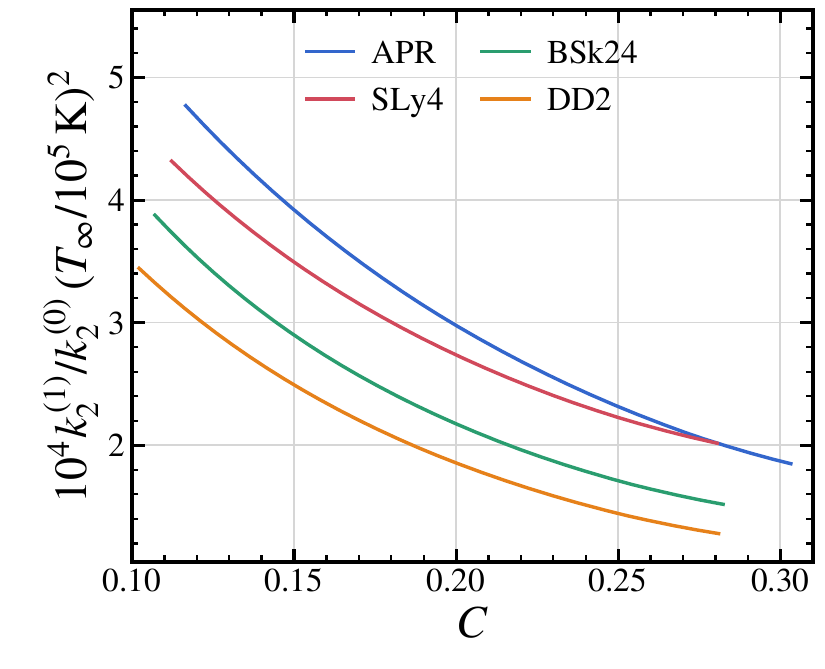} 
\caption{Linear dissipative quadrupolar Love number $\kdiss$ (left)
and ratio $\kdiss/\kstatic$ (right) as functions of compactness $C$ for
the four representative equations of state. Both panels show the
benchmark at $T_\infty=10^5\,{\rm K}$. For the viscosity law in
Eq.~\eqref{eq:viscositybenchmark}, both quantities scale as
$T_\infty^{-2}$.}
 \label{fig:publiceoslinear}
\end{figure*}

Figure~\ref{fig:publiceoslinear} shows the linear dissipative Love
number $\kdiss$ (left panel) and the ratio $\kdiss/\kstatic$ (right panel) as functions of
compactness $C$ for the four representative EOS.
Along each sequence, $\kdiss$ decreases monotonically with mass, while
the spread among the models remains modest.
At $1.4M_\odot$ and $T_\infty=10^5\,{\rm K}$, $\kdiss$ ranges from
$2.33\times10^{-5}$ (DD2) to $2.47\times10^{-5}$ (BSk24).\footnote{To compare with
Ref.~\cite{Saketh:2024juq}, we have  evaluated our entropy-production formula
using the same stellar background and viscosity profile. In particular,  we have used the same temperature
profile, rather than Tolman's isothermal condition. In this case, we have obtained
an agreement within $0.1\%$ with their
corrected result of the \texttt{v4} arXiv version.}

Because the same viscosity prescription is used throughout, the
differences among the curves arise from the EOS dependence of the
stellar structure and equilibrium tide.
The relatively narrow spread results partly from the Tolman factor
$e^{\Phi(r)}$, which suppresses the viscosity more strongly in the more
compact stellar models.
The corresponding values at $1.4M_\odot$ are reported in
Table~\ref{tab:publiceos14}.  
\begin{table}[t]
 \centering
 \begin{tabular}{@{}lcccc@{}}
 \toprule
 EOS & $R_*$ [km] & $C$ & $\kstatic$ &
 $10^5\kdiss(T_\infty/10^5\,{\rm K})^2$ \\
 \midrule
  APR   & 11.34 & 0.182 & 0.0748 & 2.45 \\
 SLy4  & 11.69 & 0.177 & 0.0775 & 2.36 \\
 BSk24 & 12.57 & 0.164 & 0.0931 & 2.47 \\
 DD2   & 13.19 & 0.157 & 0.0973 & 2.33 \\
 \bottomrule
 \end{tabular}
 \caption{Static and linear dissipative quadrupolar Love numbers for
 $m=1.4M_\odot$.  The dissipative coefficient is evaluated at fixed
 redshifted temperature $T_\infty=10^5\,{\rm K}$.}
 \label{tab:publiceos14}
\end{table}

\section{Impact on the gravitational waveform}
\label{sec:waveform}

We now compute the leading correction to the inspiral of a
quasi-circular binary due to tidal absorption.  To isolate the effect
of $\kdiss$, we use the Newtonian orbital energy and the leading
quadrupole flux, supplemented by the absorption term.  A waveform
accurate through the order of this term must also include the
point-particle corrections through that order.

The balance law for the orbital energy $U_{\rm orb}$ is
\begin{equation}
 \dot U_{\rm orb}
 =-\mathcal F_\infty-\dot m_1-\dot m_2 \;,
 \label{eq:binarybalance}
\end{equation}
where $\mathcal F_\infty$ is the gravitational-wave flux radiated to
infinity and $\dot m_a$ is the energy absorbed by body $a$.  In the
Newtonian limit, the gravitoelectric tidal field acting on body 1 is
\begin{equation}
 E_1^{ij}
 =\frac{3Gm_2}{r^3}n_o^{\langle i}n_o^{j\rangle} \;,
 \label{eq:newtonianbinarytide}
\end{equation}
where $r$ is the orbital separation, $\Omega$ is the orbital angular
frequency, and $n_o^i$ is the unit separation vector.
Using Eq.~\eqref{eq:rotatingtide} we obtain 
\begin{equation}
 \dot E_1^{ij}\dot E^1_{ij}
 =18\frac{G^2m_2^2}{r^6}\Omega^2 \;.
\end{equation}
Inserting
this equation into the EFT heating rate
\eqref{eq:EFTheating}, and using Eq.~\eqref{eq:k2definitions}, gives
\begin{equation}
 \dot m_1
 =36\lambda_{2,1}^{(1)}
   \frac{G^2m_2^2}{r^6}\Omega^2
 =6Gk_{2,1}^{(1)}R_1^6
   \frac{m_2^2}{r^6}\Omega^2 \;.
 \label{eq:binaryheating}
\end{equation}
The analogous expression for body 2 follows by interchanging the labels.

We define
\begin{equation}
 v\equiv(GM\Omega)^{1/3} \;,
 \qquad
 M\equiv m_1+m_2 \;. 
 \label{eq:binaryvariables}
\end{equation}
Using $r= {GM}/{v^2}$ and the expression of the Newtonian flux, $\mathcal F_N= {32}\nu^2v^{10}/({5G})$, where $\nu = m_1 m_2/M^2$ is the symmetric mass ratio, summing the variation of the two masses, 
Eq.~\eqref{eq:binaryheating} yields
\begin{equation} \sum_a
{\dot m_a}
 = \mathcal A_{\rm diss} v^8 {\mathcal F_N} \;,
 \label{eq:heatingfluxratio}
\end{equation}
where we have introduced, for convenience,
\begin{equation}
 \mathcal A_{\rm diss}
 \equiv
 \frac{15}{16}\sum_{a=1}^2
 \left( \frac{m_a}{M} \right)^4C_a^{-6}k_{2,a}^{(1)} \;.
 \label{eq:Adissdefinition}
\end{equation}
In terms of this, the total leading-order flux is
\begin{equation}
 \mathcal F_{\rm tot}
 =\mathcal F_N
  \left(1+\mathcal A_{\rm diss}v^8\right) \;.
 \label{eq:Adiss}
\end{equation}
The dissipative tide thus enters the flux at relative 4PN order, one PN
order before the adiabatic quadrupolar tide~\cite{Flanagan:2007ix}.
Because $\mathcal A_{\rm diss}$ contains the additional factor
$C_a^{-6}$, differences in stellar radius are amplified: although
$k_{2,a}^{(1)}$ varies by only a few percent among the models
(see Fig.~\ref{fig:publiceoslinear}),  $\mathcal A_{\rm diss}$ varies by more than a factor of two and increases
with the stellar radius, from $0.08$ for APR ($R_*=11.3\,{\rm km}$) to
$0.18$ for DD2 ($R_*=13.2\,{\rm km}$).

At this order the binding energy is
$U_N=-\nu Mv^2/2$.  Expanding the balance equation
\eqref{eq:binarybalance} to first order in $\mathcal A_{\rm diss}$ gives
\begin{equation}
 \frac{\dd t}{\dd v}
 =
 \frac{5GM}{32\nu}v^{-9}
 \left(1-\mathcal A_{\rm diss}v^8\right).
 \label{eq:timevelocity}
\end{equation}
The dissipative part of $\dd t/\dd v$ is therefore proportional to
$v^{-1}$ and integrates to $\ln v$, up to an additive 
integration constant which can be absorbed into the
coalescence time and
does not affect observables.  
Combining the resulting time and orbital phase in the
stationary-phase approximation gives (see also Ref.~\cite{Ripley:2023qxo})
\begin{equation}
 \delta\Psi_{\rm diss}(f)
 =
 -\frac{5\mathcal A_{\rm diss}}{16\nu}
 v^3\ln {v} ,
 \qquad
 v=\left(\frac{\pi GMf}{c^3}\right)^{1/3},
 \label{eq:fourierdephasing}
\end{equation}
where $f=\Omega/\pi$ is the gravitational-wave frequency and factors of
$c$ have been restored.  This has the same frequency dependence as the
4PN term
of the point-particle phase in the stationary-phase approximation~\cite{Blanchet:2023bwj}, i.e., $(3/128\nu)\,v^3\ln v\,(\psi_8+\psi_8^{\ell}\ln v)$, and is therefore
degenerate with a shift of its first coefficient,
\begin{equation}
 \delta\psi_8=-\frac{40}{3}\mathcal A_{\rm diss}
 =-\frac{25}{2}\sum_{a=1}^2\left(\frac{m_a}{M}\right)^4
 C_a^{-6}k_{2,a}^{(1)} \;.
 \label{eq:psi8shift}
\end{equation}

Since $\psi_8$ depends on the normalization of the logarithm, and the
$\psi_8^{\ell}$ term is comparable in size in the detector band, the
ratio $\delta\psi_8/\psi_8$ does not measure the size of the effect.  A
convention-independent measure is the change in the number of
gravitational-wave cycles accumulated between two frequencies.  From
$\mathcal N_{\rm GW}=\int f\,\dd t$, one obtains
(see also \cite{Saketh:2024juq})
\begin{align}
 \delta\mathcal N_{\rm GW}
 &=
 -\frac{5\mathcal A_{\rm diss}}{96\pi\nu}
 \left(v_f^3-v_i^3\right)
 \nonumber\\
 &=
 -\frac{5GM\mathcal A_{\rm diss}}
        {96\nu c^3}
 \left(f_f-f_i\right) \;,
 \label{eq:cycledephasing}
\end{align}
where $f_i$ and $f_f$ are the initial and final gravitational-wave
frequencies of the interval, with $f_i<f_f$, and $v_i=v(f_i)$ and
$v_f=v(f_f)$ are the corresponding values of the PN parameter.
The negative sign has a simple interpretation: tidal absorption increases
the total energy loss, accelerates the inspiral, and removes cycles from a
fixed frequency interval.

We evaluate these expressions for an equal-mass
$1.4+1.4M_\odot$ binary at the Tolman temperature
$T_\infty=10^5\,{\rm K}$.  For the four EOSs considered in
Sec.~\ref{sec:numerics}, the coefficient defined in
Eq.~\eqref{eq:Adissdefinition} lies in the range
\begin{equation}
 0.078\lesssim
 \mathcal A_{\rm diss}(T_\infty=10^5\,{\rm K})
 \lesssim0.18 .
 \label{eq:Adissrange}
\end{equation}
The corresponding shift of the 4PN coefficient is
$|\delta\psi_8|\simeq1.0$--$2.5$, while $\psi_8\simeq826$ and
$\psi_8^{\ell}\simeq662$ for $\nu=1/4$~\cite{Blanchet:2023bwj}.
Between $20\,{\rm Hz}$ and $1\,{\rm kHz}$, the cycle shift is
\begin{equation}
 -5.2\times10^{-4}
 \lesssim
 \delta\mathcal N_{\rm diss}
 \lesssim
 -2.2\times10^{-4}\;,
 \label{eq:equalmassdephasingrange}
\end{equation}
while the 4PN term of the time-domain point-particle
phase~\cite{Blanchet:2013haa,Marchand:2016vox,Blanchet:2023sbv,Blanchet:2023bwj}
contributes $|\delta\mathcal N_{\rm pp}^{4{\rm PN}}|\simeq0.11$ cycles
over the same interval.  Tidal dissipation thus represents
$0.2$--$0.5\%$ of the point-particle 4PN contribution at
$T_\infty=10^5\,{\rm K}$.

Extrapolating the $T_\infty^{-2}$ scaling of the dissipative correction
from Eq.~\eqref{eq:viscositybenchmark}, the dissipative and point-particle
4PN contributions to the number of cycles would become equal in magnitude
at
\begin{equation}
 T_{\infty,{\rm eq}}
 \simeq(4.5\text{--}6.9)\times10^3\,{\rm K},
 \label{eq:Teq4PNrange}
\end{equation}
and using the 4.5PN term as the reference instead changes these values by
about $5\%$.
Extending the integration formally to the Newtonian contact frequency,
\begin{equation}
 f_{\rm cont}=\frac{1}{2\pi}\left(\frac{Gm}{R_*^3}\right)^{1/2},
\end{equation}
corresponding to an orbital separation $2R_*$, i.e.,
$1.4$--$1.8\,{\rm kHz}$ for the four EOSs, the dissipative correction
remains below $1\%$ of the point-particle 4PN contribution to the number
of cycles at $T_\infty=10^5\,{\rm K}$.

The quoted equality temperatures follow from extrapolating
the benchmark scaling $\eta\propto T_\infty^{-2}$ to redshifted
temperatures below $10^4\,{\rm K}$. In this regime, pairing,
composition-dependent transport, and the applicability of a
hydrodynamic viscosity introduce substantial uncertainties.
Moreover, close to contact
the equilibrium-tide and low-frequency expansions become less controlled,
because dynamical tides and internal-mode effects can no longer be
neglected.  The robust conclusion within the regime of validity of our
calculation is that shear-viscous tidal absorption is much smaller than the
ordinary point-particle 4PN phasing.

A further limitation of the fixed-temperature estimate is that tidal
heating changes the stellar temperature during the inspiral.  Integrating
Eq.~\eqref{eq:heatingfluxratio} with Eq.~\eqref{eq:timevelocity}, while
artificially holding the temperature fixed, gives the energy absorbed by
the two stars up to a given value of $v$,
\begin{equation}
 \Delta U_{\rm abs}
 =\frac{\mathcal A_{\rm diss}\nu Mc^2}{10}v^{10}.
\end{equation}
For $T_\infty=10^5\,{\rm K}$, this reaches several $10^{47}\,{\rm erg}$ by
$1\,{\rm kHz}$, many orders of magnitude more than the thermal energy of a
neutron-star core at this temperature.

To estimate the resulting feedback on the viscosity, we follow the
temperature of the stars along the inspiral in a simple thermal model,
described in App.~\ref{app:heating}.  Starting from
$T_\infty=10^5\,{\rm K}$ well below $20\,{\rm Hz}$, the redshifted
temperature rises to $\sim10^6\,{\rm K}$ at $100\,{\rm Hz}$ and
$\sim10^7\,{\rm K}$ at $1\,{\rm kHz}$, consistent with previous estimates
of tidal heating during the inspiral~\cite{Lai:1993di,Arras:2018fxj}.  The
resulting decrease in shear viscosity reduces the magnitude of the cycle
shift between $20\,{\rm Hz}$ and $1\,{\rm kHz}$ below $10^{-6}$
(Table~\ref{tab:appheating}), with only a weak dependence on the initial temperature once
tidal heating dominates,
and a smaller heat capacity, as expected in a superfluid core, reduces it
further.  The fixed-temperature results above should therefore be
regarded as upper bounds within the benchmark transport model.

\section{Conclusion}
\label{sec:conclusions}

We have computed the leading low-frequency dissipative tidal Love number of
a nonrotating relativistic neutron star.  The central  result is Eq.~\eqref{eq:massgrowthtarget},
which connects local entropy production to the change in gravitational mass
measured at infinity.  
Matching this mass increase to the worldline heating rate reduces
$\kdiss$ to the positive radial integral \eqref{eq:k2dissintegral}.  This
representation makes passivity manifest, is linear in the shear-viscosity
profile, and requires only the standard static equilibrium-tide solution.
We  obtained the same coefficient from the imaginary,
frequency-odd exterior metric, and proved the equality of the two procedures in App.~\ref{sec:equivalence}.

For the electron shear-viscosity benchmark, the four EOS models of our main
set give $\kdiss\simeq(2.3\mbox{--}2.5)\times10^{-5}$ at $1.4M_\odot$ and
$T_\infty=10^5\,{\rm K}$.  The EOS changes this value by less than ten
percent, whereas composition, pairing, temperature, or the normalization
and radial dependence of the transport coefficient can change it
parametrically.  In particular, $\kdiss\propto T_\infty^{-2}$ decreases as
the stars heat up during the inspiral.
Our result agrees with the revised numerical results of
Ref.~\cite{Saketh:2024juq} when evaluated on the same stellar
background and with the same viscosity profile.

Shear-viscous tidal heating enters the energy balance at relative 4PN
order.  At $T_\infty=10^5\,{\rm K}$, it removes at most
$5.2\times10^{-4}$ gravitational-wave cycles between $20\,{\rm Hz}$ and
$1\,{\rm kHz}$ across the four equations of state considered here, less
than a percent of the point-particle 4PN contribution, and below $10^{-6}$
once the resulting rise in the stars' temperature is included.
For the viscosity model considered here, shear-viscous tidal
dissipation in cold neutron stars is negligible for gravitational-wave
phasing. This suggests that a measurable dissipative tidal signature
would have to come from a different mechanism, such as bulk viscosity
from weak-interaction processes~\cite{Arras:2018fxj}.

Extending the low-frequency expansion through $\ii\omega^3$ would
quantify the first correction to the linear dissipative response and
provide input for dispersion-relation constraints on compact-object
response~\cite{Apostolidis:2026mvb}. 
For slowly rotating neutron stars, the entropy-production method can
be generalized by relating the viscous heating integral to the energy
transfer in the corotating frame. This would allow us to compute
spin-dependent dissipative tidal coefficients. For black holes, the
membrane paradigm~\cite{Damour:1982surface,PriceThorne:1986,Thorne:1986membrane}
provides the corresponding horizon-shear integral, which can be used
to study the low-frequency absorptive response of Schwarzschild and
Kerr black holes. We leave these developments to future work.

\vspace{0.5cm}

{\bf Acknowledgments:}
We thank M.~V.~S.~Saketh for helpful discussions and for sharing
the revised numerical results.

\appendix

\section{Linear retarded tidal response}
\label{app:SK}

In this appendix we derive Eqs.~\eqref{eq:k2frequency} and
\eqref{eq:k2definitions} from the Schwinger--Keldysh formulation of the
worldline EFT~\cite{Galley:2012hx,Galley:2012qs}.  We restrict the discussion to the
linear quadrupolar gravitoelectric response needed in this paper.  The
generalization to the gravitomagnetic sector and to higher multipoles is
straightforward.

The tidal coupling is
\begin{equation}
 S_{\rm tid}[Q,E]
 =\int\dd\tau\,Q^{ij}(\tau)E_{ij}(\tau)\;.
 \label{eq:appinteraction}
\end{equation}
In the Schwinger--Keldysh formalism~\cite{Schwinger:1960qe,Keldysh:1964ud,Galley:2012hx} the fields are doubled, with one copy on
each branch of the closed time contour.  The starting point is the difference
of the actions on the forward and backward branches,
\begin{align}
 S_{\rm tid}^{\rm SK}
 &=S_{\rm tid}[Q_+,E_+]-S_{\rm tid}[Q_-,E_-]\;.
 \label{eq:appSKstartingpoint}
\end{align}
Introducing the combinations
\begin{align}
 Q_c^{ij}&=\frac{Q_+^{ij}+Q_-^{ij}}{2},&
 Q_r^{ij}&=Q_+^{ij}-Q_-^{ij},
 \nonumber\\
 E^c_{ij}&=\frac{E^+_{ij}+E^-_{ij}}{2},&
 E^r_{ij}&=E^+_{ij}-E^-_{ij},
 \label{eq:appkeldyshbasis}
\end{align}
Eq.~\eqref{eq:appSKstartingpoint} becomes
\begin{equation}
 S_{\rm tid}^{\rm SK}
 =\int\dd\tau\,
 \left(Q_r^{ij}E^c_{ij}+Q_c^{ij}E^r_{ij}\right).
 \label{eq:appSKinteraction}
\end{equation}
We now integrate out $Q^{ij}$.  We denote by $S_Q[Q]$ the action governing
the internal multipole degrees of freedom.  Integrating them out defines the closed-time-path generating
functional
\begin{align}
 e^{\ii W[E_c,E_r]}
 &=\int\mathcal DQ_+\mathcal DQ_-\,
 \exp\!\left\{\ii S_Q[Q_+]-\ii S_Q[Q_-]\right.
 \nonumber\\
 &\hspace{28mm}\left.
 +\ii S_{\rm tid}^{\rm SK}[Q_\pm,E_\pm]\right\}.
 \label{eq:appgeneratingfunctional}
\end{align}
The initial density matrix and the condition that the two histories are
identified at the end of the contour are left implicit.  Equivalently, the
integration may be performed over the microscopic variables from which
$Q^{ij}$ is constructed.  The detailed form of $S_Q[Q]$ is not relevant
here.

Expanding the logarithm of Eq.~\eqref{eq:appgeneratingfunctional} in
connected correlators gives
\begin{equation}
 \ii W
 =\sum_{n=1}^{\infty}\frac{\ii^n}{n!}
 \left\langle\left(S_{\rm tid}^{\rm SK}\right)^n
 \right\rangle_{Q,{\rm c}}.
 \label{eq:appcumulants}
\end{equation}
Here $\langle\cdots\rangle_{Q,{\rm c}}$ denotes a connected,
closed-time-path-ordered expectation value evaluated in the unperturbed
internal theory defined by $S_Q$.  More explicitly,
\begin{equation}
 \langle\mathcal O\rangle_Q
 \equiv
 \frac{\displaystyle
  \int\mathcal DQ_+\mathcal DQ_-\,\rho_Q\,\mathcal O\,
  e^{\ii S_Q[Q_+]-\ii S_Q[Q_-]}}
 {\displaystyle
  \int\mathcal DQ_+\mathcal DQ_-\,\rho_Q\,
  e^{\ii S_Q[Q_+]-\ii S_Q[Q_-]}},
 \label{eq:appQaverage}
\end{equation}
where $\rho_Q$ is the initial density matrix and the two histories are
identified at the final endpoint of the contour.  The subscript ${\rm c}$
in Eq.~\eqref{eq:appcumulants} denotes the connected part of this
expectation value.

The one-point function vanishes by spherical symmetry.  At quadratic order
in the tidal field, the term relevant to the mean response is the one linear
in $E_r$.  Indeed, the induced quadrupole is obtained by taking one
functional derivative with respect to $E_r$ and subsequently imposing the
physical limit $E_r=0$; terms containing two or more powers of $E_r$
therefore do not contribute.  They instead encode fluctuations and noise.
The term linear in $E_r$ is
\begin{equation}
 W_R 
 =\int\dd\tau\dd\tau'\,
 E^r_{ij}(\tau)
 G_R^{ij|kl}(\tau-\tau')
 E^c_{kl}(\tau'),
 \label{eq:appWRnonlocal}
\end{equation}
where
\begin{equation}
 G_R^{ij|kl}(\tau-\tau')
 =\ii\Theta(\tau-\tau')
 \left\langle[Q^{ij}(\tau),Q^{kl}(\tau')]\right\rangle
 \label{eq:appretardedkernel}
\end{equation}
is the retarded response kernel.  Here $Q^{ij}$ without a contour label is
the Heisenberg quadrupole operator of the physical, undoubled internal
theory; the two contour branches have combined into the retarded
commutator.  Spherical symmetry fixes the tensor
structure to
\begin{align}
 G_R^{ij|kl}(\tau)
 &=G_R(\tau)\mathcal P^{ij|kl},
 \nonumber\\
 \mathcal P^{ij|kl}
 &\equiv\frac12\left(\delta^{ik}\delta^{jl}
 +\delta^{il}\delta^{jk}\right)
 -\frac13\delta^{ij}\delta^{kl},
 \label{eq:appSTFprojector}
\end{align}
where $\mathcal P^{ij|kl}$ is the projector onto symmetric trace-free
spatial tensors.  Since $E_{ij}$ is STF, only the scalar form factor $G_R$
enters below.

The induced quadrupole follows by varying with respect to $E_r$ and then
setting $E_r=0$ and $E_c=E$,
\begin{align}
 \langle Q^{ij}(\tau)\rangle_E
 &=\left.\frac{\delta W}{\delta E^r_{ij}(\tau)}
 \right|_{E_r=0,\,E_c=E}
 \nonumber\\
 &=\int\dd\tau'\,G_R(\tau-\tau')E^{ij}(\tau').
 \label{eq:applinearresponse}
\end{align}
Here $\langle Q^{ij}\rangle_E$ denotes the expectation value of the
quadrupole in the state perturbed by the physical tidal source $E_{ij}$.
With the Fourier convention of Eq.~\eqref{eq:fourier}, this becomes
\begin{equation}
 \langle Q^{ij}(\omega)\rangle_E
 =G_R(\omega)E^{ij}(\omega).
 \label{eq:appfrequencyresponse}
\end{equation}
In the following we suppress the expectation-value brackets and the
subscript $E$ for brevity.
Causality makes $G_R(\omega)$ analytic in the upper half of the complex
frequency plane, while reality of the time-domain response implies
$G_R(-\omega)=G_R(\omega)^*$.  Its real part is therefore even in frequency
and conservative, whereas its imaginary part is odd and dissipative.  At
frequencies well below the internal-mode scales, the kernel admits the
expansion
\begin{equation}
 G_R(\omega)
 =2\left[\lambdastatic+\ii\omega\lambdadiss
 +\Order(\omega^2)\right].
 \label{eq:appGRexpansion}
\end{equation}
The factor of two is chosen such that $\lambdastatic$ is the coefficient of
$E_{ij}E^{ij}$ in the conservative worldline action.  Transforming
Eq.~\eqref{eq:appGRexpansion} back to the time domain, using
$\ii\omega\to-\partial_\tau$, gives
\begin{equation}
 W_R 
 =\int\dd\tau\,E^r_{ij}
 \left(2\lambdastatic E_c^{ij}
 -2\lambdadiss\dot E_c^{ij}\right)
 +\Order(\ddot E,E^3),
 \label{eq:appWRlocal}
\end{equation}
and hence
\begin{equation}
 Q^{ij}(\tau)
 =2\lambdastatic E^{ij}(\tau)
 -2\lambdadiss\dot E^{ij}(\tau)
 +\Order(\ddot E,E^2).
 \label{eq:appresponsetime}
\end{equation}
This local functional is the effective action $S_{\rm eff}$ of
Eq.~\eqref{eq:WRsplitQ}.

Comparing Eq.~\eqref{eq:appGRexpansion} with the definition of
$\lambda_2(\omega)$ in the main text gives Eq.~\eqref{eq:k2frequency};
extracting the conventional quadrupolar normalization then gives
Eq.~\eqref{eq:k2definitions}.  For a passive object the work done by the
dissipative quadrupole is non-negative. Equivalently, the spectral density of the retarded response
satisfies $\operatorname{Im}G_R(\omega)\geq0$ for $\omega>0$
\cite{Goldberger:2005cd,Liu:2018kfw}. Its low-frequency expansion therefore
implies $\lambda_2^{(1)}\geq0$.


\section{Metric calculation}
\label{app:metricmatching}

Here we give the details of the direct metric calculation summarized
in Sec.~\ref{sec:metricmethod}.  The derivation proceeds in three steps.
We first formulate the frequency-odd interior problem and specify its
regular central solution.  We then extract the dissipative Love number by
matching the interior solution to the static exterior basis.  Finally, we
use a Green identity to show analytically that the resulting surface
quantity is equal to the entropy-production integral
\eqref{eq:k2dissintegral}.

\subsection{Frequency-odd interior problem}
\label{app:oddinterior}

In the Fourier convention \eqref{eq:fourier}, we expand the metric
perturbations as
\begin{align}
 H_0(r,\omega)&=H(r)+\ii\omega R_*h(r)
 +\Order(\omega^2R_*^2) \;,
 \nonumber\\
 K(r,\omega)&=K(r)+\ii\omega R_*k(r)
 +\Order(\omega^2R_*^2) \;.
 \label{eq:oddmetricexpansion}
\end{align}
The static functions $H$ and $K$ are those of the equilibrium tide, while
$h$ and $k$ describe the leading response that is odd in frequency.  

At linear order in the shear viscosity, the frequency-odd functions obey
the inhomogeneous system
\begin{align}
 h'-k'+\Phi'h
 &=\frac{j_1}{R_*} \;, \label{eq:oddinteriorsystem1}
 \\
 h'+ah+bk'-\frac{2e^\Lambda}{r}k
 &=\frac{j_2}{R_*} \;,  \label{eq:oddinteriorsystem2}
\end{align}
with
\begin{equation}
\begin{aligned}
 a&=\frac{
 e^\Lambda[8+8\pi Gr^2(\bar p-\bar\rho)]-4}{2r} \;,
 \\
 b&=\frac{
 e^\Lambda(1+8\pi Gr^2\bar p)-3}{2} \;,
\end{aligned}
\end{equation}
where the sources constructed
 from the static equilibrium-tide profiles $H,K,W,V$ are given by
\begin{equation}
 j_1=\frac{16 \pi G \eta e^{-\Phi/2}}{r}
 \left(rV'+r\Phi'V-e^{\Lambda/2}W\right) \;,
 \label{eq:oddsource1}
\end{equation}
and
\begin{align}
 j_2&=\frac{4\pi G e^{-\Phi/2}}{r}
       \bigl(\eta\mathcalS_2-2r\eta' \stwo \bigr) \;,
 \label{eq:oddsource2}\\
  \mathcalS_2={}&
 -2r^2V''
 +r\left(8 \pi G r^2e^\Lambda\bar\rho-e^\Lambda+9\right)V'
 \nonumber\\
 &+\left(34e^\Lambda-6
 -16 \pi G r^2e^\Lambda\bar\rho\right)V
 \nonumber\\
 &+4e^{\Lambda/2}W+r^2e^\Lambda(H-4K) \;.
 \nonumber
\end{align}
Here $\stwo$ is the shear amplitude defined in
Eq.~\eqref{eq:shearamplitudes}.
The system \eqref{eq:oddinteriorsystem1}--\eqref{eq:oddinteriorsystem2} is obtained by inserting the
even-parity metric perturbation and the viscous stress tensor into the
linearized Einstein equations.  The trace-free angular and mixed
radial--angular equations are first used to eliminate $H_2$ and $H_1$.
The remaining two independent combinations give a first-order system
for $H_0$ and $K$.  Since every viscous contribution is already of
order $\eta\omega$, the metric and fluid functions appearing in its
source can be evaluated on the static equilibrium tide.\footnote{The
same interior system is given, in a different radial normalization and
Fourier convention, in  
Ref.~\cite{Saketh:2024juq}. }
For the adiabatically incompressible branch considered here, there is
no bulk-viscous source because the expansion of the equilibrium-tide
flow vanishes, see Eq.~\eqref{eq:vanishingexpansion}.

We next specify the regular central solution.  
If
\begin{equation}
 H=\mathcal C r^2+\Order(r^4) \;,
\end{equation}
the static equations give
\begin{align}
 K&=\mathcal C r^2+\Order(r^4) \;,
 \nonumber\\
 W&=-\frac{3\mathcal C}
 {8 \pi G(3\bar p_c+\bar\rho_c)}r^2+\Order(r^4) \;,
 \nonumber\\
 V&=\frac{3\mathcal C}
 {16 \pi G(3\bar p_c+\bar\rho_c)}r^2+\Order(r^4) \;.
 \label{eq:staticcentralseries}
\end{align}
The inhomogeneous problem is defined only up to the addition of a regular
homogeneous solution.
A convenient particular solution is selected by imposing
$h=\Order(r^4)$ at the center.  Its companion function then satisfies
\begin{equation}
 k=-\frac{6e^{-\Phi_c/2}\eta_c\mathcal C}
 {R_*(3\bar p_c+\bar\rho_c)}r^2+\Order(r^4) \;.
 \label{eq:oddcentralseries}
\end{equation}
The  freedom of adding a  homogeneous solution will drop
out of the surface quantity used below.

Integrating Eqs.~\eqref{eq:oddinteriorsystem1}--\eqref{eq:oddinteriorsystem2} with the regular central
conditions above determines $h$ and $h'$ at the stellar surface.  These
surface data give $y_1$ through Eq.~\eqref{eq:y1wronskian}, and hence the
dissipative Love number through Eq.~\eqref{eq:k2metricresult}.  We now
show that this surface extraction is identical to the
entropy-production result.

\subsection{Equivalence between the two methods}
\label{sec:equivalence}

We now prove that the metric result \eqref{eq:k2metricresult} is
identical to the entropy-production result \eqref{eq:k2dissintegral}.
The Green identity first relates the surface Wronskian to the viscous
sources in the interior. Integrating by parts then reduces those
sources to the shear kernel $\mathcal B$.

Eliminating $k'$ between the two interior equations in
App.~\ref{app:oddinterior} gives
\begin{equation}
\begin{aligned}
 k &= \frac{re^{-\Lambda}}{2}
       \bigl[(1+b)h'+(a+b\Phi')h\bigr]
       -\frac{j_\eta}{R_*},\\
 j_\eta &\equiv \frac{re^{-\Lambda}}{2}(j_2+bj_1).
\end{aligned}
\label{eq:koddalgebraic}
\end{equation}
Differentiating this expression for $k$ and substituting it back
into the first interior equation gives the same differential
operator as in the static tidal equation \eqref{eq:Hmaster}, acting
on $h$. The source is
\begin{equation}
 \frac{4e^\Lambda}{R_*r^2\Phi'}(j_\eta'-j_1),
\end{equation}
where we used the TOV relation $1+b=r\Phi'/2$.

The coefficient of the first derivative in the static tidal operator
is $2/r+(\Phi'-\Lambda')/2=\mu'/\mu$, where
$\mu\equiv r^2e^{(\Phi-\Lambda)/2}$. Multiplying the equation for $h$ by
$\mu$ therefore combines its derivative terms into $(\mu h')'$.
Multiplying the resulting equation by $H$ and subtracting the
corresponding equation for $H$ multiplied by $h$ gives
\begin{equation}
 R_*\bigl(\mu\mathcal W[H,h]\bigr)'=Z(j_\eta'-j_1) \;,
 \label{eq:greenidentitymain}
\end{equation}
where $\mathcal W[H,h]=Hh'-H'h$ is the Wronskian introduced below
Eq.~\eqref{eq:y1wronskian}, and
\begin{equation}
 Z\equiv\frac{4e^{(\Phi+\Lambda)/2}H}{\Phi'}
      =-\frac{4e^{\Phi/2}W}{r} \;.
\end{equation}
The second equality follows from Eq.~\eqref{eq:Wprofile} and the
TOV equations.

 Integrate Eq.~\eqref{eq:greenidentitymain} over a smooth radial
interval $[r_a,r_b]$. In the source integral
$\int_{r_a}^{r_b}\dd r\,Z(j_\eta'-j_1)$, integrating $Zj_\eta'$ by
parts gives the endpoint term $[Zj_\eta]_{r_a}^{r_b}$ and the bulk
term $-Z'j_\eta$. The coefficient of $\eta'$ in $j_\eta$ is
\begin{equation}
 A_1=-4\pi G r e^{-\Lambda-\Phi/2} \stwo \;.
\end{equation}
Integrating the resulting term $-Z'A_1\eta'$ by parts gives
the second endpoint term, $[-Z'\eta A_1]_{r_a}^{r_b}$.
Combining the two steps gives
\begin{equation}
\begin{split}
 &R_*\left[\mu\mathcal W[H,h]\right]_{r_a}^{r_b}\\
 &\quad=\left[Zj_\eta-Z'\eta A_1\right]_{r_a}^{r_b}\\
 &\qquad+\int_{r_a}^{r_b}\dd r\,
 \bigl[-Z'j_\eta-Zj_1+(Z'\eta A_1)'\bigr] \;.
\end{split}
\label{eq:twiceintegratedmain}
\end{equation}

The $\eta'$ terms in the integrand cancel by the definition of
$A_1$. The $V''$ terms in $j_\eta$ and $A_1'$ also cancel.
Substituting the sources and using the static tidal and displacement
equations gives
\begin{equation}
\begin{split}
 &-Z'j_\eta-Zj_1+(Z'\eta A_1)'\\
 &\qquad=-\frac{16\pi G}{3}
          \frac{\eta e^{-\Lambda/2}}{r^2}\mathcal B(r) \;.
\end{split}
\label{eq:localkernelidentity}
\end{equation}
Once the endpoint terms have been separated, the metric calculation
therefore gives the same radial dependence as the shear-heating
integral at each radius, with $\mathcal B$ given by
Eq.~\eqref{eq:positivekernel}.

At the center, regularity makes $\mu\mathcal W[H,h]$ and the viscous
endpoint terms vanish. At the surface, the viscous endpoint terms
vanish if $\eta(R_*)=\eta'(R_*)=0$. For piecewise smooth profiles,
Eq.~\eqref{eq:twiceintegratedmain} applies on each smooth interval,
with the interface terms included when matching the solutions.

To convert the surface Wronskian to the Love number, compare the
large-radius metric in Eq.~\eqref{eq:exteriordecomposition} with the
weak-field quadrupolar metric. The applied tidal potential is
$A_\uparrow r^2S/2$. A Newtonian STF mass quadrupole $I^{ij}$
therefore couples in the Lagrangian as $-I^{ij}A_\uparrow E_{ij}/2$,
so $I^{ij}=-2Q^{ij}$ in the convention of Eq.~\eqref{eq:QEcoupling}.
Its contribution to $g_{tt}$ is
\begin{equation}
 \frac{3G I^{ij}n_i n_j}{r^3}
 =-\frac{6GQ^{ij}n_i n_j}{r^3}
 =-\frac{A_\downarrow S}{r^3} \;.
\end{equation}
Using $Q^{ij}=2\lambdastatic A_\uparrow E^{ij}$ and
Eq.~\eqref{eq:k2definitions} then gives
\begin{equation}
 F(C,y_0)=\frac{A_\downarrow}{2R_*^5A_\uparrow} \;.
\end{equation}
In vacuum, $\mu\mathcal W[H_\uparrow,H_\downarrow]$ is constant
and equals $-5$ by the asymptotic normalization in
Eq.~\eqref{eq:exteriorbasis}.
Varying $A_\downarrow/A_\uparrow$ in the surface logarithmic
derivative thus gives
\begin{equation}
 F_{,y}(C,y_0)=
 -\frac{\mu(R_*)H(R_*)^2}{10R_*^6 A_\uparrow^2}.
 \label{eq:Fynormalization}
\end{equation}
Integrating Eq.~\eqref{eq:twiceintegratedmain}, taking the endpoint
terms to vanish as above, and using Eq.~\eqref{eq:y1wronskian} gives,
in the unit-tide normalization,
\begin{equation}
 \kdiss=F_{,y}(C,y_0)y_1
 =\frac{8\pi G}{15R_*^6}
  \int_0^{R_*}\dd r\,
  \frac{\eta e^{-\Lambda/2}}{r^2}\mathcal B(r).
\end{equation}
This is the entropy-production result \eqref{eq:k2dissintegral}.

\section{Heating of the stars during the inspiral}
\label{app:heating}

In this appendix we estimate how the tidal heating computed in
Sec.~\ref{sec:waveform} raises the temperature of the stars, and how the
resulting decrease of the viscosity affects the cycle shift.  
We consider an equal-mass binary and make three simplifying assumptions:
the absorbed energy is retained as heat, since cooling by neutrino and
photon emission is negligible at the temperatures reached over the
duration of the inspiral; each core maintains a uniform redshifted
temperature $T_\infty$, an idealization since the heat is deposited with
the radial profile of the shear; the redshift factors relating the
absorbed energy, measured at infinity, to the thermal energy are
neglected.

By Eq.~\eqref{eq:heatingfluxratio}, the two stars absorb the power
$\mathcal A_{\rm diss}v^8\mathcal F_N$, half of it each.  The thermal energy
$U_{\rm th}$ of each star therefore grows as
\begin{equation}
 \frac{\dd U_{\rm th}}{\dd t}=\frac12\mathcal A_{\rm diss}v^8\mathcal F_N \;.
\end{equation}
Using the leading-order term in Eq.~\eqref{eq:timevelocity},
this becomes
\begin{equation}
 \frac{\dd U_{\rm th}}{\dd v}=\frac12\mathcal A_{\rm diss}\,\nu Mc^2v^9 \;.
 \label{eq:appheatingbalance}
\end{equation}
At fixed temperature $\mathcal A_{\rm diss}$ is constant, and integrating
this equation for the two stars reproduces the absorbed energy
$\Delta U_{\rm abs}$ quoted in Sec.~\ref{sec:waveform}.

The viscosity scales as $T_\infty^{-2}$, so
$\mathcal A_{\rm diss}(T_\infty) \propto T_\infty^{-2}$. The heat capacity of degenerate matter is
proportional to temperature, $c_V\simeq10^{39}(T_\infty/10^9\,{\rm K})\,{\rm erg\,K^{-1}}$~\cite{Yakovlev:2004iq}, so that the
thermal energy is $U_{\rm th}=c_V T_\infty/2\propto T_\infty^2$.  The cycle
shift follows from $\mathcal N_{\rm GW}=\int f\,\dd t$ and the
dissipative part of Eq.~\eqref{eq:timevelocity},
\begin{equation}
 \frac{\dd\,\delta\mathcal N_{\rm diss}}{\dd v}
 =-\frac{5}{32\pi\nu}\mathcal A_{\rm diss}v^2 \;,
 \label{eq:appcycleshift}
\end{equation}
which, for constant $\mathcal A_{\rm diss}$, integrates to
Eq.~\eqref{eq:cycledephasing}.

Once the deposited heat exceeds the initial thermal energy,
Eq.~\eqref{eq:appheatingbalance} integrates to
\begin{equation}
 U_{\rm th}=\frac{1}{10}\mathcal A_{\rm diss}\,\nu Mc^2v^{10} \;,
\end{equation}
with $\mathcal A_{\rm diss}$ evaluated at the current temperature.  Since
$U_{\rm th}\propto T_\infty^2$ and
$\mathcal A_{\rm diss}\propto T_\infty^{-2}$, this gives
$T_\infty\propto v^{5/2}$ and $\mathcal A_{\rm diss}\propto v^{-5}$,
independently of the initial temperature.  Equation~\eqref{eq:appcycleshift}
then integrates to
\begin{equation}
 \delta\mathcal N_{\rm diss}\simeq
 -\frac{5}{64\pi\nu}
 \left[\mathcal A_{\rm diss}(v_i)\,v_i^3
       -\mathcal A_{\rm diss}(v_f)\,v_f^3\right],
 \label{eq:appheatedshift}
\end{equation}
with $\mathcal A_{\rm diss}$ evaluated at the temperature reached at each
endpoint.  In contrast to the fixed-temperature result, the shift is
dominated by the low-frequency end, where the stars are coldest.  Since
$\mathcal A_{\rm diss}v^8\propto v^3$, the correction also no longer takes
the 4PN logarithmic form~\eqref{eq:fourierdephasing}: its frequency dependence  mimics a 1.5PN term, with a coefficient
roughly $10^{-8}$ times that of the leading tail.

Integrating Eqs.~\eqref{eq:appheatingbalance} and~\eqref{eq:appcycleshift}
numerically from $T_\infty=10^5\,{\rm K}$ at a frequency well below
$20\,{\rm Hz}$ gives the values in Table~\ref{tab:appheating}.  The approximation~\eqref{eq:appheatedshift} reproduces the numerical
cycle shifts to better than $1\%$ for the initial conditions used here.

\begin{table}[h]
 \centering
 \begin{tabular}{@{}lcccc@{}}
 \toprule
 EOS & \multicolumn{3}{c}{$T_\infty$ [K]} &
 $\delta\mathcal N_{\rm diss}$ \\
 \cmidrule(lr){2-4}
     & $20\,{\rm Hz}$ & $100\,{\rm Hz}$ & $1\,{\rm kHz}$ &
 ($20\,{\rm Hz}$--$1\,{\rm kHz}$) \\
 \midrule
 APR   & $3.3\times10^5$ & $1.3\times10^6$ & $8.6\times10^6$ & $-5.7\times10^{-7}$ \\
 SLy4  & $3.4\times10^5$ & $1.3\times10^6$ & $9.0\times10^6$ & $-6.1\times10^{-7}$ \\
 BSk24 & $3.9\times10^5$ & $1.5\times10^6$ & $1.0\times10^7$ & $-7.8\times10^{-7}$ \\
 DD2   & $4.1\times10^5$ & $1.6\times10^6$ & $1.1\times10^7$ & $-8.7\times10^{-7}$ \\
 \bottomrule
 \end{tabular}
 \caption{Redshifted temperature and cycle shift for the equal-mass
 $1.4+1.4M_\odot$ binary when the rise of the stars' temperatures is
 included, starting from $T_\infty=10^5\,{\rm K}$ at low frequency.}
 \label{tab:appheating}
\end{table}

\bibliographystyle{utphys}
\bibliography{DLN_NS_linear}

\end{document}